\documentclass[10pt]{article}

\usepackage{amscd}
\usepackage{latexsym}
\usepackage{amsthm}
\usepackage{verbatim}
\usepackage[dutch,english]{babel}
\usepackage{amsfonts}
\usepackage{a4wide}
\usepackage[a4paper,margin=0.9in]{geometry}
\usepackage{dsfont}
\usepackage{mathtools}
\usepackage[normalem]{ulem}
\usepackage{bbm}
\usepackage{lscape}
\usepackage[T1]{fontenc} 
\usepackage{amsmath,amsbsy,amssymb,latexsym, graphics, graphicx, tabularx}
\usepackage{multirow,multicol,subfigure,booktabs}
\usepackage{algorithm}
\usepackage{adjustbox}
\usepackage{graphicx}
\usepackage{lscape}
\usepackage[noend]{algpseudocode}
\usepackage{tikz}
\usepackage{color}
\usepackage{url}
\usepackage{rotating}
\usepackage{multirow,multicol}
\usepackage{caption}
\usepackage{authblk}

\newtheorem{Theorem}{Theorem}[section] 
\newtheorem{Corollary}{Corollary}[section] 

\makeatletter

\@addtoreset{equation}{section}
\makeatother

\newcommand{\R}{\mathbb{R}}

\usepackage{cite}
\usepackage{natbib}
\setcitestyle{round,aysep={,},yysep={;}}

\usepackage{longtable}

\begin{document}
	
\title{\bf Comparison of Quantile Regression Curves with Censored Data}
\date{}
\renewcommand{\baselinestretch}{1}

\author{
{\large Lorenzo Tedesco $^*$} and
{\large Ingrid Van Keilegom} \footnote{Research financially supported by the European Research Council (2016-2022, Horizon 2020/ ERC grant agreement No. 694409).}}
\affil{ORSTAT, KU Leuven, Belgium}

\maketitle
	
\renewcommand{\baselinestretch}{1.3}

\begin{abstract} 	
This paper proposes a new test for the comparison of conditional quantile curves when the outcome of interest, typically a duration, is subject to right censoring. The test can be applied both in the case of two independent samples and for paired data, and can be used for the comparison of quantiles at a fixed quantile level, a finite set of levels or a range of quantile levels.  The asymptotic distribution of the proposed test statistics is obtained both under the null hypothesis and under local alternatives.   We describe a bootstrap procedure in order to approximate the critical values, and present the results of a simulation study, in which the performance of the tests for small and moderate sample sizes is studied and compared with the behavior of alternative tests. Finally, we apply the proposed tests on a data set concerning diabetic retinopathy.
\end{abstract}

\bigskip


\def\baselinestretch{1.2}
\newpage
\normalsize
\setcounter{footnote}{0}
\setcounter{equation}{0}
\baselineskip 18pt

\section{Introduction} \label{sec1}

Testing for the equality of distributions or quantile curves is a key task in empirical research.    Think for instance of the famous paired t-test, the two sample t-test or the Wilxocon-Mann Whitney test, which are included in any textbook on basic statistics.    In this paper, we want to study the comparison of two independent samples and the comparison of paired data by allowing for several complexities with respect to the setting of the aforementioned basic tests.   First of all, in our setting covariates are present, and we will compare the samples conditional on these covariates.  Second, we will compare the samples by means of their conditional quantiles coming from a linear quantile regression model with unknown error distribution.  And third, the response in this quantile model is allowed to be subject to right censoring.   

In the literature on the comparison of two independent samples or of paired data, several tests exist that fulfill partially the constraints of our setting, but to the best of our knowledge no paper exists so far that considers the setting we are interested in.  For instance  \cite{koul1997testing}, 
\cite{sun2006consistent},  \cite{dette2011comparing}, \cite{dette2013nonparametric}, 
\cite{delgado2013conditional}, 
 among others, are examples of papers in which conditional quantile functions are compared.  However, these papers assume that the response is fully observed, which is often not the case in experiments involving duration variables.   On the other hand, in \cite{li1996quantile} a comparison of quantile curves in the case of censored data is presented, but their setup does not include covariates.  The inclusion of covariates in the comparison of quantile curves, could lead to significant advances in this area.  Finally, \cite{vilar2004nonparametric} present a nonparametric comparison of curves with dependent errors, but their procedure does not work for quantile curves or censored data, and the dependency is intended to be present among the observations in each sample, and not between the samples.

Another important contribution in this area is the paper by \cite{pardo2006comparison}, who consider the comparison of regression curves in the case of censored outcomes.   However, the location functional in their regression model belongs to a certain class of L-functionals, which includes certain truncated means as special case, but it does not include any quantiles. Moreover, it is assumed that the error in their model is independent of the covariates, which we do not want to assume.   So again, the framework is different from the one we want to study.  The paper by \cite{lee2009nonparametric} presents a method to compare conditional distributions in case of censored responses, which is again a related but different research question than the one we want to study.  The covariates in this paper do not necessarily have the same distribution in the two samples, which will also be the case in our setting, but the comparison of distribution functions is different from the comparison of quantiles for a single or finite set of quantile levels.  Also note that the comparison of quantities like the conditional median, allows to obtain specific insights in the relative behavior of the two samples.   A final somewhat related paper is the one by \cite{sant2014nonparametric}, who provides a method to compare conditional distributions  and single quantile curves in the case of censored responses. However, this method requires the covariates in the two samples to have the same distribution, which could be problematic in certain practical situations.  Moreover, the regression model in this paper is fully nonparametric, whereas we prefer to work with linear regression models, which are easier to interpret and which do not suffer from the curse-of-dimensionality problem.

The paper is organized as follows. In Section \ref{sec2} we present the quantile regression model and introduce the null hypothesis we like to test.  Section \ref{sec3} presents the test statistics for the comparison both in the case of paired data and in the case of independent samples, and the asymptotic distribution of the proposed tests is shown under high-level conditions on the estimators of the conditional quantiles, both under the null hypothesis and under local alternatives. In Section \ref{sec4}  these high-level conditions are verified for a number of specific estimators.   In Section \ref{sec5} a bootstrap procedure is presented to approximate the critical values of the tests. In Section \ref{sec6} the performance of the proposed tests is studied by means of simulations, whereas in Section \ref{sec7} data on diabetic retinopathy
are analyzed using the proposed methodology.  
Finally, Section \ref{sec8} concludes the paper and discusses ideas for future research, while the Appendix contains the proofs of the main results and additional simulation results.

\section{The model} \label{sec2}

We consider a general framework for the comparison of two quantile regression curves, that are estimated either based on paired data or on independent samples.  The framework allows for the comparison of quantile curves at one quantile level, a finite set of levels, or an interval of quantile levels.   We focus on the case where the response variable in the two samples is subject to random right censoring, but the adaptation to other types of incomplete data can be handled similarly.   Moreover,  the proposed approach is based on generic estimators of the quantile curves in the two samples.  The estimators need to satisfy certain high-level conditions, that can be satisfied for specific estimators.  

We need to introduce the following notations. For $j=1,2$, the survival time $T_j > 0$ satisfies
\begin{align} \label{model}
\log T_j = \beta_{j0}(\tau) + \beta_{j1}(\tau) Z_{j,1} + \ldots + \beta_{jp}(\tau) Z_{j,p} + \epsilon_j(\tau). 
\end{align}
Here $Z_j=(1,Z_{j,1},\ldots,Z_{j,p})^T$ is a vector of covariates and the distribution of the error $\epsilon_j(\tau)$ is left unspecified except that $P(\epsilon_j(\tau) \le 0 | Z_j) = \tau$.   So the $\tau$-th quantile of the error $\epsilon_j(\tau)$ given $Z_j$ equals 0.   For a given $\tau \in (0,1)$, let $\beta_j(\tau) = (\beta_{j0}(\tau),\beta_{j1}(\tau),\ldots,\beta_{jp}(\tau))^T$ be the vector of coefficients, so that we can also write $\log T_j = Z_j^T \beta_j(\tau) + \epsilon_j(\tau)$. The $\log$-transformation of the time variable is not strictly required, but it is common to transform the (positive) response $T_j$ to a variable defined on the whole real line, to obtain a better fit in practice.

We are interested in testing 
\begin{align} \label{H0}
H_0 : \beta_1(\tau) = \beta_2(\tau) \quad \mbox{for all } \tau \in A, 
\end{align}
versus the general alternative  under which there is at least one value of $\tau$ in $A$ for which $\beta_1(\tau)$ is not equal to $\beta_2(\tau)$.  Here, $A$ is either a singleton, a finite set of points, or an interval in $(0,1)$.  Note that adaptations of the null hypothesis $H_0$ can be considered.  For instance, it is possible to test whether $R\beta_1(\tau) = R\beta_2(\tau)$ with $R$ a matrix, or whether $\gamma(\beta_1(\tau)) = \gamma(\beta_2(\tau))$, where $\gamma$ is a possibly non-linear but differentiable function. This implies, in particular, the possibility of testing $\beta_1(\tau)^T z  = \beta_2(\tau)^T z$ for a fixed covariate vector $z$, i.e.\ we can test whether there is a significant difference between two quantiles for specific values of the covariates.    We will not elaborate further on this, but it is useful to know that this is possible.

Because of random right censoring, instead of observing $T_j$ for $j=1,2$ we observe $X_j = \min(T_j,C_j)$ and $\Delta_j = I(T_j \le C_j)$, where the censoring time $C_j$ is supposed to be independent of $T_j$ given $Z_j$.   The data consist of two i.i.d.\ samples $(X_{ji},\Delta_{ji},Z_{ji})_{i=1}^{n_j}$ with the same distribution as $(X_j,\Delta_j,Z_j)$ for $j=1,2$, where $n_1$ and $n_2$ are the sample sizes in case of independent samples, and $n_1=n_2=n$ in the case of paired data.  In the latter case, the data are collected from the same subjects (for instance for two eyes or at two different time points) and hence $(X_{1i},\Delta_{1i},Z_{1i})$ and $(X_{2i},\Delta_{2i},Z_{2i})$ are dependent for each fixed $i$.  

Based on these data, we will estimate the vectors of regression coefficients $\beta_1(\tau)$ and $\beta_2(\tau)$.   The literature on linear quantile regression with censored data contains a rich collection of papers, for instance \cite{portnoy2003censored}, \cite{peng2008survival}, \cite{wang2009locally}, \cite{portnoy2010asymptotics}, \cite{yang2018new}, \cite{debacker2019} and \cite{de2020linear}. 
For more details about each of these estimators, we refer the reader to the recent review by \cite{Peng2021}.  In this paper we will consider the estimators proposed by \cite{portnoy2003censored}, \cite{peng2008survival}, \cite{wang2009locally} and \cite{debacker2019}. 
The last two estimators can be used for a single value of $\tau$, whereas the estimator of \cite{portnoy2003censored} and \cite{peng2008survival}   satisfy certain asymptotic results uniformly over a range of $\tau$-values.    

We finish this section with a remark regarding the case of paired data.  Our null hypothesis implies that the difference of the quantiles equals zero, but in the case of paired data it could also be of interest to test whether the quantile of the difference between two paired data points equals zero.  Unlike the case of the mean function, the quantile function is not linear (in the sense that the quantile of the difference is not the same as the difference of the quantiles) and hence these two quantities are not equal. 
We believe that the difference of the quantiles is more meaningful in our context, since it is the quantity that allows for a general comparison of distributions. 

\section{The test and asymptotic theory} \label{sec3}

In this section we will propose test statistics for testing $H_0$, and we will develop the asymptotic distributions of these test statistics separately for the two settings (paired samples or independent samples).  The test statistics will be based on certain estimators $\hat\beta_1(\cdot)$ and $\hat\beta_2(\cdot)$ for the two samples.  These estimators depend on the actual estimation method that is used.  In this section we will therefore work with generic estimators that satisfy certain regularity conditions.  In the next section these regularity conditions will be verified for specific estimators.   

Define $n=n_1+n_2$ in the case of independent samples, whereas in the case of paired samples $n$ represents the number of paired data.    The proposed test statistics depend on a weight function $w(\tau)$ and are given for the case where the set $A$ is an interval.   If $A$ is a finite set of points (or contains just one quantile level), the integrals below need to be replaced by sums over the elements of $A$.  We consider three test statistics:  
\begin{itemize}
\item Averaged $L_2$-norm: 
\begin{eqnarray} \label{L2}
T_{L_2} = n^{1/2} \int_A \|\hat\beta_1(\tau)-\hat\beta_2(\tau)\|_2 w(\tau) \, d\tau, 
\end{eqnarray}
where $\|\cdot\|_2$ is the Euclidean norm in $\R^{p+1}$.   
\item Averaged $L_\infty$-norm:
\begin{eqnarray} \label{Linfty}
T_{L_\infty} = n^{1/2} \int_A \|\hat\beta_1(\tau)-\hat\beta_2(\tau)\|_{\infty} w(\tau) \, d\tau, 
\end{eqnarray}
where $\|\cdot\|_\infty$ is the supremum norm in $\R^{p+1}$.  
\item Averaged absolute value with Bonferroni correction:
\begin{eqnarray} \label{Bonf}
T_{B,k} = n^{1/2} \int_A |\hat\beta_{1k}(\tau)-\hat\beta_{2k}(\tau)| w(\tau) \, d\tau 
\end{eqnarray}
for $k=0,\ldots,p$, which consists in rejecting $H_0$ if at least one of the p-values corresponding to the test statistics $T_{B,0},\ldots,T_{B,p}$ is smaller than $\alpha/(p+1)$, where $\alpha$ is the chosen significance level. 
\end{itemize}

Note that other test statistics could have been considered (like for instance the supremum norm \linebreak $n^{1/2} \sup_\tau \|\hat\beta_1(\tau)-\hat\beta_2(\tau)\|_\infty$), but we opted for the ones presented above because of better simulation results.  Also note that the test statistics can be adapted in an obvious way to the case where one is only interested in testing whether one of the components of $\beta_1$ equals the corresponding component of $\beta_2$.   This could be of interest if only the effect of one specific covariate is of interest.

We now consider the asymptotic theory for these test statistics separately for paired samples and independent samples.   For both cases, we suppose that the estimators $\hat\beta_1(\cdot)$ and $\hat\beta_2(\cdot)$ can be written as a sum of i.i.d.\ terms plus a remainder term of negligible asymptotic order.  The verification of this condition for specific estimators will be done in the next section.

\begin{itemize}
\item[(A)] The estimator $\hat\beta_j(\cdot)$ ($j=1,2$) satisfies the following i.i.d.\ representation for all $\tau \in A$ and $k=0,\ldots,p$:
$$ \hat\beta_{jk}(\tau) - \beta_{jk}(\tau) = n_j^{-1} \sum_{i=1}^{n_j} g_{jk}(\tau,X_{ji},\Delta_{ji},Z_{ji}) + R_{jk}(\tau), $$ 
for some function $g_{jk}$ satisfying $E[g_{jk}(\tau,X_j,\Delta_j,Z_j)|Z_j]=0$ depending on the estimator $\hat\beta_j(\tau)$, where $\sup_{\tau \in A}|R_{jk}(\tau)| = o_P(n_j^{-1/2})$. 
Moreover, the class ${\cal F}_{jk} = \{(x,\delta,z) \rightarrow g_{jk}(\tau,x,\delta,z) : \tau \in A\}$ is $P_j$-Donsker, where $P_j$ is the joint probability law of $(X_j,\Delta_j,Z_j)$.
\end{itemize}

The representation in condition (A) will in general not hold for all $0<\tau<1$.  Due to the right censoring mechanism, the estimation of $\beta_j(\tau)$ will not be possible for $\tau$ larger than a certain threshold value.   See Section \ref{sec4} for more details.  Hence, the choice of the set $A$ depends on this threshold value.

Let $g_j=(g_{j0},\ldots,g_{jp})$ for $j=1,2$.   As we will see in the next section, this condition holds for all the estimators mentioned in Section \ref{sec2} for a single value of $\tau$, and also for an interval of $\tau-$values in case of the estimator of \cite{peng2008survival} and \cite{portnoy2003censored}.

\subsection{Independent samples}

In what follows, we will work under the following equal growth (EG) assumption on the sample sizes $n_1$ and $n_2$:

\begin{itemize}
\item[(EG)] $n_1/n \rightarrow p_1$ and $n_2/n \rightarrow p_2$ as $n$ tends to infinity, where $0<p_1,p_2<1$.
\end{itemize}

We then have the following result.

\begin{Theorem} \label{theo1}
Assume (A) and (EG).  Then, under $H_0$, the $(p+1)$-dimensional process $n^{1/2}(\hat\beta_1(\tau)-\hat\beta_2(\tau))$ indexed by $\tau \in A$ converges weakly to a zero-mean Gaussian process $W_I$ with covariance function
\begin{align*}
Cov(W_I(\tau_1),W_I(\tau_2)) = \, & p_1^{-1} E \Big[g_1(\tau_1,X_1,\Delta_1,Z_1) g_1(\tau_2,X_1,\Delta_1,Z_1)^T\Big], \\
& + p_2^{-1} E \Big[g_2(\tau_1,X_2,\Delta_2,Z_2) g_2(\tau_2,X_2,\Delta_2,Z_2)^T\Big],
\end{align*}
for $\tau_1,\tau_2 \in A$. 
\end{Theorem}
The proof is given in the Appendix.\\ 
Note that, in case $A$ is a finite set of points instead of an interval, the Gaussian process reduces to a multivariate normal random vector. Based on this result and the continuous mapping theorem, we have the following corollary regarding the asymptotic distribution of the test statistics. 

\begin{Corollary} \label{cor1}
Assume (A) and (EG).  Then, under $H_0$,
\begin{align*}
& T_{L_2} \stackrel{d}{\rightarrow} \int_A \|W_I(\tau)\|_2 w(\tau) \, d\tau, \quad\quad  T_{L_\infty} \stackrel{d}{\rightarrow} \int_A \|W_I(\tau)\|_\infty w(\tau) \, d\tau, \\
& T_{B,k} \stackrel{d}{\rightarrow} \int_A |W_{I,k}(\tau)| w(\tau) \, d\tau,
\end{align*}
for $k=0,\ldots,p$. 
\end{Corollary}

As before, the above result is developed for the case where $A$ is an interval.  If $A$ is a finite set of points it suffices to replace the integrals by sums.  

Consider now the limiting behaviour of the test statistics under the following local alternative:
$$ H_1 : \beta_1(\tau) - \beta_2(\tau) = n^{-1/2} b(\tau) \quad \mbox{for all } \tau \in A, $$
for some continuous and integrable function $b$.
Then, we have the following result: 

\begin{Corollary} \label{cor2}
Assume (A) and (EG).  Then, under $H_1$,
\begin{align*}
& T_{L_2} \stackrel{d}{\rightarrow} \int_A \|W_I(\tau) + b(\tau)\|_2 w(\tau) \, d\tau, \quad\quad T_{L_\infty} \stackrel{d}{\rightarrow} \int_A \|W_I(\tau) + b(\tau)\|_\infty w(\tau) \, d\tau, \\
& T_{B,k} \stackrel{d}{\rightarrow} \int_A |W_{I,k}(\tau) + b_k(\tau)| w(\tau) \, d\tau,
\end{align*}
for $k=0,\ldots,p$. 
\end{Corollary}
The proof is given in the Appendix. \\

\subsection{Paired samples}

The results in the case of paired samples can be obtained in a similar way as for independent samples.  The major difference lies in the derivation of the covariance function of the limiting process, as is shown in the next result. 

\begin{Theorem} \label{theo2}
Assume (A).  Then, under $H_0$, the $(p+1)$-dimensional process $n^{1/2}(\hat\beta_1(\tau)-\hat\beta_2(\tau))$ indexed by $\tau \in A$ converges weakly to a zero-mean Gaussian process $W_P$ with covariance function
\begin{align*}
Cov(W_P(\tau_1),W_P(\tau_2)) = \, & E \Big[\big\{g_1(\tau_1,X_1,\Delta_1,Z_1) - g_2(\tau_1,X_2,\Delta_2,Z_2)\big\} \\
& \times \big\{g_1(\tau_2,X_1,\Delta_1,Z_1) - g_2(\tau_2,X_2,\Delta_2,Z_2)\big\}^T\Big],
\end{align*}
for $\tau_1,\tau_2 \in A$. 
\end{Theorem}
The proof is given in the Appendix. \\
Similarly as for independent samples we now obtain the following corollary, in which we combine the results under $H_0$ and $H_1$ (it suffices to take $b \equiv 0$ under $H_0$):

\begin{Corollary} \label{cor3}
Assume (A).  Then, under $H_1$,
\begin{align*}
& T_{L_2} \stackrel{d}{\rightarrow} \int_A \|W_P(\tau) + b(\tau)\|_2 w(\tau) \, d\tau, \quad\quad T_{L_\infty} \stackrel{d}{\rightarrow} \int_A \|W_P(\tau) + b(\tau)\|_\infty w(\tau) \, d\tau, \\
& T_{B,k} \stackrel{d}{\rightarrow} \int_A |W_{P,k}(\tau) + b_k(\tau)| w(\tau) \, d\tau,
\end{align*}
for $k=0,\ldots,p$. 
\end{Corollary}

\section{Choice of the quantile regression estimator} \label{sec4}

In this section we will present some of the quantile estimators that have been proposed in the literature when the response is censored.   We will explain how these estimators are defined, and for which choices of the set $A$ and the functions $g_j$ ($j=1,2$) assumption (A) is satisfied for these estimators.   This will then allow us to conclude that the asymptotic results shown in the previous section are valid for these estimators.

\subsection{Estimator of \cite{peng2008survival}}\label{sec:estPeng}

The estimator utilises a martingale structure underlying randomly censored data to construct an estimating equation for the model in (\ref{model}). In particular, let $\Lambda_T(t|Z) = -\log(1-P(T\leq t|Z))$ be the cumulative hazard function of $T$ given $Z$, $N(t) =I(X \le t, \Delta = 1)$ and $M(t)=N(t) - \Lambda_T(t \wedge X|Z) $. Let $N_{ji}(t)$ and $M_{ji}(t)$ be the analogue of $N(t)$ and $M(t)$ for observation $i$ of sample $j=1,2$. Then, $M_{ji}(t)$ is the martingale process associated with the counting process $N_{ji}(t)$. So, $E[M_{ji}(t)|Z_{ji}] = 0$ for $t>0$ and 
$$ E \Big\{\sum_{i=1}^{n_j} Z_{ji} \Big[N_{ji}(\exp\big(Z_{ji}^T\beta_j(\tau))\big) - \Lambda_T \big(\exp(Z_{ji}^T\beta_j(\tau)) \wedge X_{ji} | Z_{ji} \big) \Big] \Big\} = 0 $$ 
for each $\tau$.  Now, $Z_{ji}^T\beta_j(\tau)$ is monotone in $\tau$, and therefore it can be easily seen that $\Lambda_T(\exp(Z_{ji}^T\beta_j(\tau))\wedge X_{ji}|Z_{ji}) = \int_0^\tau I(X_{ji}\geq \exp(Z_{ji}^T\beta_j(u)))dH(u)$,  where $H(x) = -\log(1-x)$, provided model (\ref{model}) holds for all $0 \le u \le \tau$. These results motivate to solve the estimating equation 
\begin{equation}
\label{eq:pengHuangEstimation}
    S_{jn_j}(\beta_j,\tau)=0,
\end{equation} where
\begin{equation} 
S_{jn_j}(\beta_j,\tau) = \frac{1}{n_j} \sum_{i=1}^{n_j} Z_{ji} \Big[ N_{ji}(\text{exp} \big(Z_{ji}^T \beta_j(\tau))\big) - \int_0^{\tau} I \big(X_{ji} \geq \text{exp}(Z_{ji}^T\beta_j(u))\big) dH(u) \Big].
\end{equation}
An estimator $\hat\beta_j(\tau)$ of $\beta_j(\tau)$ can therefore be obtained by approximating the stochastic solution of (\ref{eq:pengHuangEstimation}). Under mild conditions  (see Section 3 in \cite{peng2008survival}) the estimator is uniformly consistent and converges weakly to a Gaussian process.

What remains to prove is that assumption (A) is satisfied for $A=[\tau_L,\tau_R]$, where $0<\tau_L<\tau_R$ and $\tau_R$ has to be smaller than the cumulative distribution of $T_j$ at the upper bound of the support of $C_j$ for $j=1,2$.  As shown in the proof of Theorem 2 in \cite{peng2008survival}, the class
\begin{equation}  \label{eq:donskePengHuang}
\mathcal{C} = \Big\{Z_{ji} \Big[N_{ji}( \text{exp}\big(Z_{ji}^T \beta_j(\tau))\big) - \int_0^{\tau} I\big(X_{ji} \geq \text{exp}(Z_{ji}^T\beta_j(u))\big) dH(u)\Big] : \tau \in [\tau_L, \tau_R] \Big\}
\end{equation} 
is Donsker with $\beta_j(\tau)$ equal to the true parameter, and moreover,
\begin{equation}
\label{eq:Acondition1}
E \Big\{Z_{ji} \Big[N_{ji}\big( \text{exp}(Z_{ji}^T \beta_j(\tau))\big) - \int_0^{\tau} I \big(X_{ji} \geq \text{exp}(Z_{ji}^T\beta_j(u))\big) dH(u) \Big] \Big| Z_{ji} \Big\} = 0
\end{equation}
by martingale properties. Therefore, the quantity $S_{jn_j}$ can be seen as an empirical process evaluated over the Donsker class (\ref{eq:donskePengHuang}).  Now, it is proved in \cite{peng2008survival} that the difference $n^{1/2}_j(\hat{\beta_j}(\tau) - \beta_j(\tau))$ can be written as 
\begin{equation}
\label{eq:Acondition2}
    n^{1/2}_j(\hat{\beta_j}(\tau) - \beta_j(\tau)) = n^{1/2}_j \tilde{\phi}(-S_{jn_j}(\beta_j,\tau)) +  o_{[\tau_L,\tau_R]}(1),
\end{equation}
where $o_I(1)$ is a quantity that converges uniformly to zero in probability for $\tau \in I$ and $\tilde{\phi}$ is a linear map, and so it preserves the Donsker property ($\tilde{\phi}(\cdot)(\tau)$ is the map $B(\beta(\tau))^{-1}\phi(\cdot)(\tau)$ defined on page 648 of \cite{peng2008survival}). Note that (\ref{eq:donskePengHuang})-(\ref{eq:Acondition2}) are equivalent to saying that $\hat{\beta}_j$ satisfies condition (A), where $g_{jk}$ corresponds to the $k$-th component of $g_j(\tau,X_{j},\Delta_{j},Z_{j}) = Z_{j} \big[N_{j}( \text{exp}\big(Z_{j}^T \beta_j(\tau))\big) - \int_0^{\tau} I\big(X_{j} \geq \text{exp}(Z_{j}^T\beta_j(u))\big) dH(u)\big]$. 

\subsection{Other Estimators}
In this section we briefly discuss three other estimators available in the literature, namely the estimators proposed by \cite{wang2009locally}, \cite{portnoy2003censored} and \cite{debacker2019} and we explain why condition (A) is satisfied for these estimators.

\subsubsection{Estimator of \cite{wang2009locally} }
The estimator proposed in \cite{wang2009locally} is based on a modification of the standard quantile loss function by twisting the idea of the self-consistent Kaplan-Meier estimator.   This means that the estimator is constructed by redistributing to the right the probability mass associated with each censored data point by means of a local weighting scheme. Now, the asymptotic distribution of $n_j^{1/2}(\hat\beta_j(\tau) - \beta_j(\tau))$ is obtained by showing that this quantity is a sum of i.i.d.\ random vectors with mean zero, plus an error term of order $o_P(1)$, and then applying the central limit theorem. Therefore, for a single value of $\tau$, $n_j^{1/2}(\hat\beta_j(\tau) - \beta_j(\tau))$ can be seen as an empirical process corresponding to a family composed of a single function, and so it is Donsker. In conclusion, also the \cite{wang2009locally} estimator satisfies condition (A) when the set $A$ is a singleton.   This estimator does not require that model (\ref{model}) is valid for all quantile levels smaller than $\tau$.

\subsubsection{Estimator of \cite{portnoy2003censored}}
By adapting the idea of redistributing censoring probabilities 
in the self-consistent Kaplan-Meier estimator, the estimator of \cite{portnoy2003censored} uses an iterative self-consistent algorithm to estimate the globally linear quantile regression model.  
The algorithm proposed in the original paper was simplified into a grid-based sequential estimation procedure by \cite{neocleous2006correction}, and the corresponding asymptotic study is reported in \cite{portnoy2010asymptotics}.
In the latter paper it is shown that $n_j^{1/2}(\hat\beta_j(\tau) - \beta_j(\tau))$ converges to a normal random variable for fixed $\tau$.
\cite{peng2012note} proposed an alternative formulation of the self-consistent approach based on stochastic integral equations, and showed that the estimator (say $\hat\beta_{P,j}(\tau)$) is asymptotically equivalent to the \cite{peng2008survival} estimator (say $\hat\beta_{PH,j}(\tau)$) in the sense that $\sup_{\tau\in [\tau_L,\tau_R]} \| {n_j^{1/2}} (\hat\beta_{P,j}(\tau) - \hat\beta_{PH,j}(\tau) )\|_2  \rightarrow_p 0$.  
In conclusion, to show the validity of condition (A) for the estimator of \cite{portnoy2003censored}, an argument as for \cite{wang2009locally} can be used in case the set $A$ is a singleton. For multiple values of $\tau$, it is not clear how to show condition (A) in a direct way, but the equivalence with the \cite{peng2008survival} estimator can be used, since we have shown condition (A) for the latter estimator. 
\subsubsection{Estimator of \cite{debacker2019}}
 Another possible estimator is proposed in \cite{debacker2019}. The estimator is based on an adjusted standard quantile loss function which  accommodates randomly censored data. More precisely, define $ G_{C|Z}(c|Z) = \mathbb{P}(C\ge c|Z)$, then the derivative of $ \phi_{\tau}(t, Y, G_{C|Z}(\cdot|Z)) = (Y-t)\{\tau - I(Y\le t)\} - (1-\tau)\int_0^t\{1-G_{C|Z}(s|Z)\}ds$ with respect to $t$, is equal to $-\{ I(Y>t)-G_{C|Z}(t|Z)(1-\tau)\}$. So, conditional on $Z$, the derivative has expectation zero when $ t = \beta(\tau)^\top Z$ under model \eqref{model}. In the aforementioned paper, this property is used to construct an adjusted loss function for censored quantile regression, namely $\sum_{i=1}^{n}\phi_\tau(Z_i^\top\beta, Y_i, \hat G_{C|Z}(\cdot|Z_i))$, with $\hat G_{C|Z}(\cdot|Z_i)$ a consistent estimator of the survival function of $C$ given $Z$.  Depending on the dependence between $C$ and $T$, different choices for $\hat G_{C|Z}$ are possible and we refer to the original paper and to \cite{Peng2021} for further details. Then, by minimizing the non-convex adjusted loss function it is possible to obtain a consistent and asymptotically normal estimator $\hat\beta(\tau)$ of $\beta(\tau)$. \cite{debacker2019} also provide a numerically robust MM algorithm for such minimization. The validity of condition (A), when the set $A$ is a singleton, is thanks to Theorems 3.1 and 3.2 in \cite{debacker2019}.

\section{Bootstrap procedure} \label{sec5}


We will now develop a general bootstrap approach that can be used to approximate the critical values of the test statistics described in Section \ref{sec2}, avoiding in this way the estimation of unknown density functions that appear in the expression of the limiting process, or other functions that are difficult to estimate. The use of bootstrap  techniques to approximate the distribution of test statistics in the context of comparison of curves has already been explored in the literature, see for instance \cite{vilar2007bootstrap}. The general approach will be discussed in relation with the quantile estimators available in the literature. 

Suppose that we have a bootstrap estimator $\hat \beta_j^*(\cdot)$ that satisfies the following condition:

\begin{itemize}
\item[(A$^*$)] The bootstrap estimator $\hat\beta_j^*(\cdot)$ ($j=1,2$) satisfies the following i.i.d.\ representation for all $\tau \in A$ and $k=0,\ldots,p$: 
$$ \hat\beta^*_{jk}(\tau) -  \hat\beta_{jk}(\tau) = n_j^{-1} \sum_{i=1}^{n_j} \eta_{ji} g_{jk}(\tau,X_{ji},\Delta_{ji},Z_{ji}) + R^*_{jk}(\tau), $$ 
for some i.i.d.\ variables $\eta_{ji}$, $i=1, \ldots, n_j, j=1,2$, with mean zero and variance equal to 1 in the case of independent samples, and for some i.i.d.\ variables $\eta_{1i}=\eta_{2i}$, $i=1, \ldots, n=n_1=n_2$, with mean zero and variance equal to 1 in the case of paired samples.  Here, $\sup_{\tau \in A}|R^*_{jk}(\tau)| = o_{P^*_j}(n_j^{-1/2})$ 
 for almost all realizations of the data $(X_{ji},\Delta_{ji},Z_{ji})_{i=1}^{n_j}$, the functions $g_{jk}$ are the same as in condition (A),  and $P^*_j$  stands for the probability measure of the $\eta_{ji}$-variables.
\end{itemize}

Based on this, we will show that we can use the empirical percentiles of $\hat\beta_1^*(\tau) - \hat\beta_2^*(\tau) - \hat\beta_1(\tau) + \hat\beta_2(\tau)$ to approximate those of $\hat\beta_1(\tau) - \hat\beta_2(\tau) - \beta_1(\tau) + \beta_2(\tau)$. Therefore, it will be possible to test $H_0$ by comparing the bootstrap process with $\hat\beta_1(\tau) - \hat\beta_2(\tau)$.   The following theorem holds both for the case of independent samples and paired samples (in which case $n_1=n_2=n$). 

\begin{Theorem} \label{theo3}

Assume (A$^*$), and assume the hypotheses of Theorem \ref{theo1} in the case of independent samples and those of Theorem \ref{theo2} in the case of paired samples. Then, $n^{1/2}(\hat\beta_1 -\hat\beta_2 - \beta_1 + \beta_2)(\tau)$ converges to a Gaussian process $W(\tau)$ indexed by $\tau \in A$, and $n^{1/2}(\hat\beta_1^* - \hat\beta_2^* - \hat\beta_1 + \hat\beta_2)  (\tau)$ converges to the same Gaussian process  $W(\tau)$ for $\tau \in A$ conditionally on the data, a.s.

\end{Theorem}

The proof is given in the Appendix.   


\subsection{Estimator of \cite{peng2008survival}} \label{sec5a}
To show the validity of condition (A$^*$) for the \cite{peng2008survival} estimator, we define
\begin{equation}
    \label{eq:pengHuangEstimationBootstrap}
    S_{jn_j}^*(\beta,\tau) = \frac{1}{n_j} \sum_{i=1}^{n_j} \tilde\eta_{ji} Z_{ji} \Big[ N_{ji}\big(\text{exp}(Z_{ji}^T \beta(\tau))\big) - \int_0^{\tau} I \big(X_{ji} \geq \text{exp}(Z_{ji}^T\beta(u))\big) dH(u) \Big], 
\end{equation}
where $\tilde\eta_{ji}$ ($i=1,\ldots,n_j$, $j=1,2$) are i.i.d.\ with a unit exponential distribution in the case of independent samples, and $\tilde\eta_{1i}=\tilde\eta_{2i}$ ($i=1,\ldots,n$) are i.i.d.\ with a unit exponential distribution in the case of paired samples.   \cite{peng2008survival}  define the bootstrap estimator $\hat\beta^*_j(\tau)$ as an approximate solution of the equation $S_{jn_j}^*(\beta,\tau)=0$ in $\beta$. They show in Appendix C that $\hat\beta^*_j(\tau)$ satisfies 
\begin{align} 
n_j^{1/2}S_{jn_j}^*(\hat\beta_j^*,\tau) &= o_{[\tau_L,\tau_R]}(1) \,\,a.s. \label{eq:PHC1} \\
n_j^{1/2}\{ \hat\beta_j^*(\tau) - \hat\beta_j(\tau)\} &= \tilde\phi(-n_j^{1/2}S_{jn_j}^*(\hat\beta_j,\tau)) + o_{[\tau_L,\tau_R]}(1), \label{eq:PHC2}
\end{align}
where again $o_{[\tau_L,\tau_R]}(1)$ is a quantity converging to zero in probability, uniformly in $[\tau_L,\tau_R]$,  and $\tilde\phi$ is the same linear map that appears in \eqref{eq:Acondition2}. 
 
We showed in Section \ref{sec:estPeng} that the $k$-th component of $\tilde\phi(-n_j^{1/2}S_{jn_j}(\hat\beta_j,\tau))$ is equivalent to the sum that appears on the right hand side of condition (A), and moreover we specified the functions $g_{jk}$. Since $\tilde \phi $ is linear, it is easy to see that the $k$-th component of the quantity $\tilde\phi(-n_j^{1/2}S_{jn_j}^*(\hat\beta_j,\tau))$ corresponds to the $k$-th component of $\tilde\phi(-n_j^{1/2}S_{jn_j}(\hat\beta_j,\tau))$ with $g_{jk}$ replaced by $\eta_{ji} g_{jk}$ and $\eta_{ji} = \tilde \eta_{ji}-1$. Hence, condition (A$^*$) follows.
The implementation of this re-weighting bootstrap procedure is described in Section 4.1 of \cite{peng2008survival}.

\subsection{Other estimators} \label{sec5b}

In the papers by \cite{wang2009locally}, \cite{portnoy2003censored},  and \cite{debacker2019} the asymptotic validity of a bootstrap procedure has not been shown.   As suggested in these papers, we will work with a naive bootstrap method, that consists in the case of independent samples of resampling with replacement the triplets $\{(X_{ji},\Delta_{ji},Z_{ji})\}_{i=1}^{n_j}$ separately for $j=1$ and $j=2$, whereas in the case of paired samples we draw resamples with replacement from the joint sample $\{(X_{1i},\Delta_{1i},Z_{1i},X_{2i},\Delta_{2i},Z_{2i})\}_{i=1}^{n}$. Note that this bootstrap procedure can be justified by classical bootstrap theory and we refer to Section 3.3 of \cite{wang2009locally}, Section 6.1 of \cite{portnoy2003censored} and Section 3 of \cite{debacker2019} for more information.

\section{Simulations} \label{sec6}

In this section we report the results of simulations in order to evaluate the finite sample performance of the proposed test statistics for testing $H_0$. For both the cases of independent and paired samples, we test the equality of a single quantile curve, specifically the median, and the equality over a range of quantile levels. We study the performance of the test statistics described in Section \ref{sec3} and of the estimators discussed in Section \ref{sec4}.

For testing the equality of the medians the set $A=\{0.5\}$ is a singleton, and so any of the four estimators described in Section \ref{sec4} can be used.  
Instead, for testing the equality over a range of quantile levels (or testing the equality of the corresponding distributions over a range of time points), only the estimator of \cite{peng2008survival} and \cite{portnoy2003censored} can be used, since the validity of the asymptotic results for a set of $\tau$-values is only shown for these two estimators. However, we decide to also include the results of the \cite{wang2009locally} and \cite{debacker2019} estimators in order to compare the test performance.

The proposed tests consist in accepting or rejecting the null hypothesis based on the bootstrap approach presented in Section \ref{sec5}. Specifically, let $t_a^*$ be the quantile of level $0<a<1$ of the bootstrap test statistic, for any of the test statistics defined in Section \ref{sec3}.  Then, the test consists in rejecting the null hypothesis if the test statistic for the given sample is outside the interval $[t_{\alpha/2}^*,t_{1-\alpha/2}^*]$.   However, the performance of the test statistics can be improved by standardizing the process $n^{1/2}(\hat\beta_1-\hat\beta_2)$ and the corresponding bootstrap process.  More precisely, we will work with $n^{1/2}\hat\Sigma(\tau)^{-1/2}(\hat\beta_1-\hat\beta_2)(\tau)$ , where  $\hat\Sigma(\tau)$ is a bootstrap estimator of the covariance matrix.  The corresponding bootstrap process is $n^{1/2}\hat\Sigma(\tau)^{-1/2}[(\hat\beta_1^*-\hat\beta_1)(\tau)-(\hat\beta_2^*-\hat\beta_2)(\tau)]$.   The proposed standardization does not change the asymptotic properties of the test, but turns out to be particularly appropriate in order to increase the power. This can be explained observing that, when the components of $\beta_1-\beta_2$ have different scales, the differences in the smaller components are not observable when the norm of the non-standardized process is considered. 

In order to optimize the computation time of the simulation studies, we approximate the rejection probabilities using the method proposed in \cite{giacomini2013warp}.  For a study with (say) $N$ simulation runs, the method works as follows. For each of the $N$ samples we compute the process  $\hat\beta_1(\tau) - \hat\beta_2(\tau)$ and we compute the  corresponding bootstrap version $\hat\beta_1^*(\tau) - \hat\beta_2^*(\tau) - \hat\beta_1(\tau) + \hat\beta_2(\tau)$ for a single bootstrap sample drawn from that given sample. All the $N$ simulations contribute in the estimation of the rejection probability. We refer to the aforementioned paper for further details.

\subsection{Independent samples}

We first consider the case of independent samples, taking into account three different models reported in Table \ref{table:models}.  The error $\epsilon_j$ follows a normal distribution with mean zero and variance 0.25 ($N(0,0.25)$), 
which is independent of the covariate vector $Z_j=(Z_{j,1},Z_{j,2})$, $j=1,2$. Note that Model 1 can also be written as in equation \eqref{model}, where $\beta_{j0}(\tau) = \beta_{j0}+\Phi^{-1}(\tau) 0.5$, $\beta_{j1}(\tau) = \beta_{j1}$, $\beta_{j2}(\tau) = \beta_{j2}$ and $\epsilon_j(\tau) = \epsilon_j - \Phi^{-1}(\tau) 0.5$, where $\Phi^{-1}(\tau)$ is the quantile function of the standard normal distribution.  In this way, the error $\epsilon_j(\tau) $ satisfies $P(\epsilon_j(\tau) \le 0 | Z_j) = \tau$ as is required in (\ref{model}).   In a similar way, Model 3 in Table \ref{table:models} can be written in the form of (\ref{model}). Finally, for Model 2 we have that $\beta_{j0}(\tau) = \beta_{j0}+\beta_{j2} \Phi^{-1}(\tau) 0.5$, $\beta_{j1}(\tau) = \beta_{j1}$, $\beta_{j2}(\tau) = \Phi^{-1}(\tau)0.5$ and $\epsilon_j(\tau) = (\beta_{j2} + Z_{j,2}) (\epsilon_j - \Phi^{-1}(\tau)0.5)$, which satisfies $P(\epsilon_j(\tau) \le 0 | Z_j) = \tau$ provided $\beta_{j2} + Z_{j,2}>0$, since $\epsilon_j$ is independent of $Z_{j,2}$.  

\begin{table}[H]
\centering 
\begin{tabular}{|c|l|}
\hline
Model 1 & $\log T_j = \beta_{j0} + \beta_{j1} Z_{j,1} + \beta_{j2}  Z_{j,2}  + \epsilon_j$ \\ \hline
Model 2 & $\log T_j = \beta_{j0} + \beta_{j1 }Z_{j,1} + (\beta_{j2} + Z_{j,2}) \epsilon_j$ \\ \hline
Model 3 & $\log T_j = \beta_{j0} + \beta_{j1} Z_{j,1} + \beta_{j2}  Z_{j,2}  + \frac{1}{2}\epsilon_j$ \\ \hline
\end{tabular}
\caption{\label{table:models} Specification of the models used in the simulations ($j=1,2$).}
\end{table}

For each model the coefficients $\beta_{10}=\beta_{20}=0,\beta_{11}=-0.5, \beta_{12}=\beta_{22}=0.5$ are fixed, whereas the value of $\beta_{21}$ varies, namely $\beta_{21} = -0.5, -0.3$ or $-0.1$.  So the null hypothesis $H_0: \beta_1(\tau)=\beta_2(\tau)$ will be satisfied when $\beta_{21}=-0.5$ and fails in the two other cases.  Moreover, we fix the size of the first sample to $n_1=200$ and vary the second sample size by taking $n_2=100, 200, 400$. The censoring time $C_j$ is uniformly distributed on the interval $[0,c_j]$ and is independent of $(T_j,Z_j)$, where $c_j$ is chosen in order to have on average 20\% and 40\% of censored data in each sample $j=1,2$. 

As already noticed, our method does not require the distribution of the covariates to be equal in the two samples.   In order to allow comparisons with competing methods, we will include simulations for both equally distributed and not equally distributed covariates.  

In the first setting, the covariates have the same distribution in the two samples: $Z_{j,1}$ has a uniform distribution $U[0,1]$, and $Z_{j,2}$ has a Bernoulli distribution $B(0.5)$ with success parameter 0.5, for each $j=1,2$. The covariates $Z_{1,1}, Z_{1,2}, Z_{2,1}$ and $Z_{2,2}$ are mutually independent.  In this setting, we make a comparison of distributions, choosing as set $A$ the interval $[\tau_L,\tau_R] = [0.1,0.6]$, 
and step size equal to $0.01$.  
We compare the performance of our method with the one proposed in  \cite{sant2014nonparametric}. In that paper, a nonparametric method is developed for the comparison of two conditional distributions corresponding to two independent samples with equally distributed covariates. 
The method depends on a propensity score $p(\cdot)$, which we estimate with a series logit estimator using a power series of order 1 for $n_2=100$, of order 3 for $n_2=200$ and of order 5 for $n_2=400$, as suggested in Section 5 of \cite{sant2014nonparametric}. 
Since the method of \cite{sant2014nonparametric} does not assume that the quantiles are linear in the covariates, some loss of power can be expected with respect to our procedure.

In the second setting the covariates have different distributions in the two samples: $Z_{1,1} \sim U[0,1], Z_{2,1} \sim U[0,1.2]$, $Z_{1,2} \sim B(0.5)$ and $Z_{2,2} \sim B(0.7)$.  In this case we compare the quantiles of the two samples for $\tau=0.5$, corresponding to the median, under the three models given above.   To the best of our knowledge, no competitor exists in the case of independent samples with unequal covariate distributions. A summary of the two simulation settings can be found in Table \ref{table:settingSpecification}.
\begin{table}[H]
\centering
\begin{tabular}{|c|c|c|c|c|c|}
\hline
Setting & $Z_{1,1}$  & $Z_{1,2}$ & $Z_{2,1}$    & $Z_{2,2}$ & Performed comparison        \\ \hline
1       & $U[0,1]$ & $B(0.5)$    & $U[0,1]$   & $B(0.5)$    & Distribution ($\tau \in [0.1,0.6]$)     \\ \hline
2       & $U[0,1]$ & $B(0.5)$    & $U[0,1.2]$ & $B(0.7)$    & Quantile function ($\tau = 0.5$) \\ \hline
\end{tabular}
\caption{\label{table:settingSpecification} Specification of the distribution of  the covariates $Z_{i,j}$ for $i = 1,2$, $j=1,2$ in Settings 1 and 2 together with the performed comparison.  Here, $B(p)$ and $U[a,b]$ indicate the Bernoulli distribution with parameter $p$ and the uniform distribution on the interval $[a,b]$, respectively.    }
\end{table}

For both settings, we use the estimators discussed in Section \ref{sec4}, that are  \cite{debacker2019},  \cite{peng2008survival}, \cite{portnoy2003censored} and \cite{wang2009locally}, and  the test statistics $T_{L_2}$ , $T_{L_\infty}$ and $T_{B,k}$ given in (\ref{L2}), (\ref{Linfty}) and (\ref{Bonf}).   We repeat the simulations 500 times for each model, and calculate the rejection probabilities based on a significance level of $\alpha=0.05$ (and $\alpha=0.05/3$ for the Bonferroni correction).  The critical values of the tests are calculated based on the weighted bootstrap method described in Section  \ref{sec5a} for the method of \cite{peng2008survival}, and a naive bootstrap method with replacement for the other estimators (see Section \ref{sec5b}).  The method of \cite{sant2014nonparametric} is based on a multiplier-type bootstrap procedure, that is somewhat similar to the bootstrap described in Section \ref{sec5a}.  The number of bootstrap iterations for all bootstrap methods is $N = 500$.

The simulation results for Setting 1 are reported in Table \ref{table:independentDistribution20} and, included in the Appendix, in Table \ref{table:independentDistribution40},  for the case of 20\% and 40\% of censored data, respectively. The proposed method is able to detect the equality of the distributions, with the following specifications. When the $H_0$ hypothesis holds, the type I error for the estimator of \cite{peng2008survival} is close to the theoretical value of 5\%.  This is not always true for the estimators of \cite{portnoy2003censored} and \cite{wang2009locally}.
Interestingly, the results of the estimator of \cite{debacker2019} are comparable with those of \cite{peng2008survival}.
Among the proposed test statistics (Bonferroni, $T_{L_2}$ and $T_{L_\infty}$), the results are comparable when the alternative hypothesis holds. However, the Bonferroni statistic has a higher rejection rate under $H_0$. Also note that the power of the tests correctly increases when one of the following occurs: the sample size $n_2$ increases, the difference $\beta_{21} - \beta_{11}$ increases, or the percentage of censored data decreases. The method of \cite{sant2014nonparametric} obtains comparable or worse results for all cases.  This can be explained by the fact that that method does not use the assumption of linearity of the quantiles, which is used by our method. As Model 3 is equal to Model 1 except that the error term has smaller variance, it is interesting to compare the results.  We notice that under Model 3 the rejection rates under the alternative hypothesis ($\beta_{21} - \beta_{11} = 0.2$ or 0.4) are larger than under Model 1, as can be expected.  

\begin{table}[H]
\centering
\begin{adjustbox}{width=1\textwidth}
\small
\begin{tabular}{|c|c|c|ccc|ccc|ccc|ccc|c|} 
\cline{4-16}
\multicolumn{3}{c|}{\multirow{2}{*}{}} & Bonf. & $T_{L_2}$ & $T_{L_\infty}$ & Bonf. & $T_{L_2}$ & $T_{L_\infty}$ & Bonf. & $T_{L_2}$ & $T_{L_\infty}$ & Bonf. & $T_{L_2}$ & $T_{L_\infty}$ &        \\ 
\hline
Mod. & $n_2$ & Diff.     & \multicolumn{3}{c|}{\cite{debacker2019}}   & \multicolumn{3}{c|}{\cite{peng2008survival}} & \multicolumn{3}{c|}{\cite{portnoy2003censored}}   & \multicolumn{3}{c|}{\cite{wang2009locally}}       & SA     \\ 
\hline
1    & 100   & 0                       & 0.070 & 0.040     & 0.042        & 0.040 & 0.022   & 0.024        & 0.012 & 0.004   & 0.010        & 0.024 & 0.010        & 0.006        & 0.066  \\
     &       & 0.2                     & 0.202 & 0.150     & 0.160        & 0.144 & 0.108   & 0.128        & 0.018 & 0.032   & 0.032        & 0.197 & 0.165        & 0.133        & 0.148  \\
     &       & 0.4                     & 0.636 & 0.608     & 0.548        & 0.654 & 0.608   & 0.572        & 0.462 & 0.520   & 0.470        & 0.597 & 0.647        & 0.565        & 0.516  \\ 
\cline{2-16}
     & 200   & 0                       & 0.052 & 0.024     & 0.034        & 0.054 & 0.032   & 0.034        & 0.076 & 0.034   & 0.034        & 0.050 & 0.034        & 0.030        & 0.032  \\
     &       & 0.2                     & 0.282 & 0.290     & 0.278        & 0.300 & 0.280   & 0.236        & 0.214 & 0.228   & 0.216        & 0.322 & 0.314        & 0.294        & 0.254  \\
     &       & 0.4                     & 0.836 & 0.868     & 0.854        & 0.860 & 0.902   & 0.858        & 0.870 & 0.918   & 0.880        & 0.864 & 0.898        & 0.866        & 0.702  \\ 
\cline{2-16}
     & 400   & 0                       & 0.046 & 0.040     & 0.036        & 0.034 & 0.026   & 0.032        & 0.030 & 0.018   & 0.020        & 0.056 & 0.038        & 0.036        & 0.022  \\
     &       & 0.2                     & 0.350 & 0.382     & 0.344        & 0.406 & 0.380   & 0.380        & 0.198 & 0.230   & 0.206        & 0.372 & 0.406        & 0.374        & 0.358  \\
     &       & 0.4                     & 0.902 & 0.936     & 0.922        & 0.926 & 0.960   & 0.934        & 0.860 & 0.890   & 0.858        & 0.902 & 0.958        & 0.916        & 0.792  \\ 
\hline
2    & 100   & 0                       & 0.032 & 0.022     & 0.018        & 0.040 & 0.028   & 0.034        & 0.018 & 0.006   & 0.010        & 0.016 & 0.008        & 0.016        & 0.032  \\
     &       & 0.2                     & 0.370 & 0.402     & 0.346        & 0.288 & 0.280   & 0.232        & 0.204 & 0.214   & 0.162        & 0.253 & 0.257        & 0.204        & 0.106  \\
     &       & 0.4                     & 0.938 & 0.956     & 0.948        & 0.918 & 0.950   & 0.912        & 0.820 & 0.912   & 0.840        & 0.948 & 0.954        & 0.908        & 0.556  \\ 
\cline{2-16}
     & 200   & 0                       & 0.052 & 0.036     & 0.028        & 0.052 & 0.054   & 0.046        & 0.030 & 0.024   & 0.026        & 0.046 & 0.022        & 0.026        & 0.050  \\
     &       & 0.2                     & 0.474 & 0.498     & 0.430        & 0.546 & 0.546   & 0.524        & 0.570 & 0.572   & 0.520        & 0.630 & 0.524        & 0.500        & 0.222  \\
     &       & 0.4                     & 0.998 & 0.998     & 0.998        & 0.990 & 0.990   & 0.986        & 0.998 & 0.998   & 0.996        & 0.998 & 1.000        & 1.000        & 0.752  \\ 
\cline{2-16}
     & 400   & 0                       & 0.032 & 0.020     & 0.020        & 0.056 & 0.024   & 0.040        & 0.026 & 0.012   & 0.018        & 0.072 & 0.038        & 0.042        & 0.050  \\
     &       & 0.2                     & 0.660 & 0.714     & 0.666        & 0.678 & 0.730   & 0.676        & 0.506 & 0.582   & 0.512        & 0.654 & 0.656        & 0.620        & 0.350  \\
     &       & 0.4                     & 1.000 & 1.000     & 1.000        & 1.000 & 1.000   & 1.000        & 1.000 & 1.000   & 0.998        & 1.000 & 1.000        & 1.000        & 0.844  \\ 
\hline
3    & 100   & 0                       & 0.048 & 0.034     & 0.034        & 0.030 & 0.012   & 0.016        & 0.016 & 0.000   & 0.000        & 0.028 & 0.014        & 0.018        & 0.048  \\
     &       & 0.2                     & 0.606 & 0.600     & 0.580        & 0.574 & 0.666   & 0.604        & 0.358 & 0.386   & 0.316        & 0.677 & 0.649        & 0.629        & 0.162  \\
     &       & 0.4                     & 0.996 & 0.998     & 0.996        & 0.998 & 1.000   & 0.998        & 1.000 & 1.000   & 1.000        & 1.000 & 1.000        & 1.000        & 0.752  \\ 
\cline{2-16}
     & 200   & 0                       & 0.034 & 0.032     & 0.030        & 0.030 & 0.034   & 0.024        & 0.058 & 0.042   & 0.056        & 0.046 & 0.014        & 0.022        & 0.046  \\
     &       & 0.2                     & 0.860 & 0.866     & 0.852        & 0.830 & 0.872   & 0.830        & 0.796 & 0.858   & 0.804        & 0.844 & 0.870        & 0.830        & 0.306  \\
     &       & 0.4                     & 1.000 & 1.000     & 1.000        & 1.000 & 1.000   & 1.000        & 1.000 & 1.000   & 1.000        & 1.000 & 1.000        & 1.000        & 0.934  \\ 
\cline{2-16}
     & 400   & 0                       & 0.036 & 0.022     & 0.026        & 0.036 & 0.026   & 0.026        & 0.010 & 0.006   & 0.008        & 0.010 & 0.008        & 0.006        & 0.040  \\
     &       & 0.2                     & 0.958 & 0.966     & 0.942        & 0.944 & 0.964   & 0.942        & 0.834 & 0.890   & 0.808        & 0.936 & 0.938        & 0.924        & 0.490  \\
     &       & 0.4                     & 1.000 & 1.000     & 1.000        & 1.000 & 1.000   & 1.000        & 1.000 & 1.000   & 1.000        & 1.000 & 1.000        & 1.000        & 0.970  \\
\hline
\end{tabular}
\end{adjustbox}
\caption{\label{table:independentDistribution20} Simulation results for the comparison of quantile curves (range of $\tau$-values) for two independent samples for percentage of censoring equal to 20\%.  The size of the first sample is $n_1=200$, the significance level is $\alpha=0.05$ and Diff $=\beta_{21}-\beta_{11}$, so the null hypothesis is satisfied for Diff $=0$. The abbreviation SA stands for the test statistic of \cite{sant2014nonparametric}, and Bonf, $T_{L_2}$ and $T_{L_\infty}$ stand for the test statistics defined in (\ref{Bonf}), (\ref{L2}) and (\ref{Linfty}), respectively. }
\end{table}

The results of Setting 2 are reported in Table  \ref{table:independentMedian20} and, included in the Appendix, in Table \ref{table:independentMedian40}, for 20\% and 40\% of censored data, respectively. 
When $H_0$ holds, the type I error for the estimator of \cite{peng2008survival}, \cite{debacker2019} and \cite{wang2009locally} is close to the theoretical value of 5\%. Instead, for the  \cite{portnoy2003censored} estimator, we again obtain a worse result. 
The fact that the type I error is not exactly equal to $0.05$ also for the case $n_2 = 400$, can be explained by the fact that the value of $n_1$ remains fixed ($n_1=200$). We also simulated the case where $n_1,n_2$ both increase, and in that case all the test statistics reached the theoretical type I error. Again, the power of the tests correctly increases when the sample size $n_2$ increases, the difference $\beta_{21} - \beta_{11}$ increases, or the percentage of censored data decreases.
Finally, we note that, not surprisingly, the power of the tests considering a range of $\tau$-values is higher than the power for a single quantile curve. This can be observed by comparing Table \ref{table:independentDistribution20} and Table \ref{table:independentMedian20} (although the distributions of the covariates are not exactly the same, the large difference in p-values gives quite convincing evidence). 


\begin{table}[H]
\centering
\begin{adjustbox}{width=1\textwidth}
\small
\begin{tabular}{|c|c|c|ccc|ccc|ccc|ccc|}
\multicolumn{1}{c}{} & \multicolumn{1}{c}{} & \multicolumn{1}{c}{} &       &           & \multicolumn{1}{c}{} &       &           & \multicolumn{1}{c}{} &       &           & \multicolumn{1}{c}{} &       &         & \multicolumn{1}{c}{}  \\ 
\cline{4-15}
\multicolumn{1}{c}{} & \multicolumn{1}{c}{} &                      & Bonf  & $T_{L_2}$ & $T_{L_\infty}$       & Bonf  & $T_{L_2}$ & $T_{L_\infty}$       & Bonf  & $T_{L_2}$ & $T_{L_\infty}$       & Bonf  & $T_{L_2}$ & $T_{L_\infty}$          \\ 
\hline
Mod.                 & $n_2$                & Diff.                & \multicolumn{3}{c|}{De Backer}           & \multicolumn{3}{c|}{PengHuang}           & \multicolumn{3}{c|}{Portnoy}             & \multicolumn{3}{c|}{WangWang}           \\ 
\hline
1                    & 100                  & 0                    & 0.006 & 0.006     & 0.006                & 0.016 & 0.006     & 0.012                & 0.006 & 0.002     & 0.008                & 0.008 & 0.014   & 0.010                 \\
                     &                      & 0.2                  & 0.064 & 0.096     & 0.072                & 0.072 & 0.056     & 0.056                & 0.018 & 0.032     & 0.016                & 0.072 & 0.070   & 0.082                 \\
                     &                      & 0.4                  & 0.304 & 0.388     & 0.282                & 0.424 & 0.422     & 0.394                & 0.128 & 0.302     & 0.178                & 0.330 & 0.460   & 0.350                 \\ 
\cline{2-15}
                     & 200                  & 0                    & 0.016 & 0.012     & 0.020                & 0.022 & 0.010     & 0.020                & 0.042 & 0.042     & 0.038                & 0.006 & 0.006   & 0.004                 \\
                     &                      & 0.2                  & 0.098 & 0.124     & 0.086                & 0.168 & 0.224     & 0.186                & 0.108 & 0.188     & 0.122                & 0.112 & 0.130   & 0.108                 \\
                     &                      & 0.4                  & 0.596 & 0.712     & 0.614                & 0.576 & 0.626     & 0.596                & 0.738 & 0.774     & 0.720                & 0.686 & 0.766   & 0.714                 \\ 
\cline{2-15}
                     & 400                  & 0                    & 0.038 & 0.030     & 0.040                & 0.024 & 0.012     & 0.028                & 0.002 & 0.006     & 0.002                & 0.028 & 0.016   & 0.024                 \\
                     &                      & 0.2                  & 0.246 & 0.266     & 0.242                & 0.180 & 0.228     & 0.182                & 0.096 & 0.110     & 0.092                & 0.260 & 0.280   & 0.258                 \\
                     &                      & 0.4                  & 0.794 & 0.870     & 0.816                & 0.830 & 0.878     & 0.794                & 0.718 & 0.768     & 0.684                & 0.684 & 0.784   & 0.688                 \\ 
\hline
2                    & 100                  & 0                    & 0.028 & 0.026     & 0.022                & 0.010 & 0.006     & 0.010                & 0.014 & 0.016     & 0.010                & 0.012 & 0.004   & 0.010                 \\
                     &                      & 0.2                  & 0.112 & 0.150     & 0.142                & 0.134 & 0.198     & 0.138                & 0.062 & 0.098     & 0.072                & 0.124 & 0.156   & 0.142                 \\
                     &                      & 0.4                  & 0.734 & 0.832     & 0.740                & 0.784 & 0.786     & 0.738                & 0.614 & 0.752     & 0.588                & 0.651 & 0.782   & 0.685                 \\ 
\cline{2-15}
                     & 200                  & 0                    & 0.028 & 0.026     & 0.024                & 0.032 & 0.038     & 0.026                & 0.038 & 0.028     & 0.038                & 0.020 & 0.014   & 0.016                 \\
                     &                      & 0.2                  & 0.322 & 0.374     & 0.318                & 0.372 & 0.376     & 0.336                & 0.406 & 0.448     & 0.422                & 0.372 & 0.438   & 0.348                 \\
                     &                      & 0.4                  & 0.956 & 0.978     & 0.946                & 0.942 & 0.960     & 0.942                & 0.946 & 0.976     & 0.948                & 0.954 & 0.976   & 0.944                 \\ 
\cline{2-15}
                     & 400                  & 0                    & 0.020 & 0.026     & 0.022                & 0.016 & 0.020     & 0.022                & 0.006 & 0.002     & 0.006                & 0.006 & 0.006   & 0.006                 \\
                     &                      & 0.2                  & 0.558 & 0.606     & 0.552                & 0.292 & 0.322     & 0.266                & 0.308 & 0.414     & 0.318                & 0.460 & 0.560   & 0.528                 \\
                     &                      & 0.4                  & 0.996 & 0.998     & 0.998                & 1.000 & 1.000     & 1.000                & 0.980 & 0.994     & 0.980                & 0.996 & 0.998   & 0.996                 \\ 
\hline
3                    & 100                  & 0                    & 0.020 & 0.012     & 0.022                & 0.014 & 0.018     & 0.016                & 0.006 & 0.000     & 0.006                & 0.004 & 0.002   & 0.008                 \\
                     &                      & 0.2                  & 0.282 & 0.426     & 0.324                & 0.330 & 0.394     & 0.326                & 0.266 & 0.308     & 0.230                & 0.334 & 0.424   & 0.364                 \\
                     &                      & 0.4                  & 0.984 & 0.992     & 0.984                & 0.988 & 0.994     & 0.988                & 0.960 & 0.982     & 0.960                & 0.978 & 0.990   & 0.980                 \\ 
\cline{2-15}
                     & 200                  & 0                    & 0.026 & 0.022     & 0.026                & 0.008 & 0.002     & 0.008                & 0.026 & 0.016     & 0.022                & 0.030 & 0.022   & 0.024                 \\
                     &                      & 0.2                  & 0.676 & 0.774     & 0.660                & 0.598 & 0.688     & 0.582                & 0.636 & 0.688     & 0.638                & 0.652 & 0.752   & 0.646                 \\
                     &                      & 0.4                  & 0.996 & 1.000     & 0.998                & 1.000 & 1.000     & 1.000                & 1.000 & 1.000     & 1.000                & 1.000 & 1.000   & 1.000                 \\ 
\cline{2-15}
                     & 400                  & 0                    & 0.024 & 0.020     & 0.024                & 0.020 & 0.020     & 0.022                & 0.004 & 0.000     & 0.004                & 0.042 & 0.024   & 0.030                 \\
                     &                      & 0.2                  & 0.838 & 0.842     & 0.818                & 0.792 & 0.878     & 0.804                & 0.648 & 0.752     & 0.668                & 0.748 & 0.772   & 0.734                 \\
                     &                      & 0.4                  & 1.000 & 1.000     & 1.000                & 1.000 & 1.000     & 1.000                & 1.000 & 1.000     & 1.000                & 1.000 & 1.000   & 1.000                 \\
\hline
\end{tabular}
\end{adjustbox}
\caption{\label{table:independentMedian20} Simulation results for the comparison of the median ($\tau=0.5$) for two independent samples subject to $20\%$ censoring. The size of the first sample is $n_1=200$, the significance level is $\alpha=0.05$ and Diff $=\beta_{21}-\beta_{11}$, so the null hypothesis is satisfied for Diff $=0$. The abbreviations Bonf, $T_{L_2}$ and $T_{L_\infty}$ stand for the test statistics defined in (\ref{Bonf}), (\ref{L2}) and (\ref{Linfty}), respectively.  }
\end{table}


\subsection{Paired samples}
We report here the simulation results in the case of paired samples, where the three different models specified in Table \ref{table:models} are considered. As before, for each model the coefficients $\beta_{10}=\beta_{20}=0,\beta_{11}=-0.5, \beta_{12}=\beta_{22}=0.5$ are fixed, whereas the value of $\beta_{21}$ varies, namely $\beta_{21} = -0.5, -0.3$ or $-0.1$. The covariates have the same values among the samples, and are distributed as $Z_{1,1} = Z_{2,1} \sim U[0,1]$ and $Z_{1,2} =Z_{2,2} \sim B(0.5)$. The errors $\epsilon_1$ and $\epsilon_2$ for the two samples are now dependent and have a bivariate normal distribution, specifically \begin{align}\label{eq:dependent}
    \Big(\begin{array}{c} \epsilon_1 \\ \epsilon_2 \end{array} \Big) \sim N_2 \Big(\Big(\begin{array}{c} 0 \\ 0 \end{array} \Big), \Big(\begin{array}{cc} 0.5 & \eta \\ \eta & 0.5 \end{array} \Big) \Big), \end{align} 
where $\eta = 0.2, 0.4$ denotes the covariance between $\epsilon_1$ and $\epsilon_2$. Note that, for $j = 1,2$, the marginal distribution  of $\epsilon_j$ is $N(0,0.5)$. Therefore, with the same arguments as in the previous section, we can express each model as in (\ref{model}).

The sample size $n_1 = n_2 = n$ is equal to $n =100, 200, 400$. The censoring time $C_j$ is equal among the two samples  (so, $C_1 = C_2 = C$), and $C$ has as uniform distribution on the interval $[0,c]$, where $c$ is chosen in order to obtain on average 20\% and 40\% of censored data in the first sample. Note that equal censoring times among the two samples are not required by the proposed method, but we opt for this choice since it is often the case, in empirical experiments, that the censoring occurs simultaneously for paired individuals. The same argument holds for the chosen covariate values $Z_{1,j} = Z_{2,j}$ for $j=1,2$.   

As before we include the results regarding the comparison of distributions and single quantile curves ($\tau = 0.5$).  In this case it was not possible to compare our method with alternative ones, since, as far as we know, no competitor exists for the comparison of quantile curves based on paired censored data.  We include the results for the test statistics presented in Section \ref{sec3} and the estimators discussed in Section \ref{sec4}. We repeat the simulations 500 times for each model, and calculate the rejection probabilities based on a significance level of $\alpha=0.05$ (and $\alpha=0.05/3$ for the Bonferroni correction).  The critical values of the tests are calculated based on the paired weighted bootstrap method described in Section  \ref{sec5a} for the method of \cite{peng2008survival}, and a paired naive bootstrap method with replacement for the other estimators (see Section \ref{sec5b}). For the distribution comparison, the set $A$ is again the interval $[\tau_L,\tau_R] = [0.1,0.6]$ with step size equal to $0.01$. 
 
The results for the comparison of distributions for paired samples are reported in Table \ref{table:pairedDistribution20} and, included in the Appendix, in Table \ref{table:pairedDistribution40}, for 20\% and 40\%  of censored data, respectively.  The type I error is  close to the theoretical value of 5\% for the estimators of \cite{peng2008survival} and \cite{portnoy2003censored}, which satisfy condition (A) also when $A$ is different from a singleton, but the first obtains, in general, better results. On the other hand, the estimators of  \cite{wang2009locally} and \cite{debacker2019} obtain a worse type I error rate when $n=400$.  Note that when the sample size $n$ or the difference $\beta_{21} - \beta_{11}$ increases, or the percentage of censored data decreases, the power of the test correctly increases. This occurs also when the covariance between the errors increases, as can be expected.  In general, the test is conservative: the type I error rate is close to the theoretical value of 5\% only for large sample sizes. This shows the difficulty of the comparison of distributions in case of dependent data.

The results for the comparison of a single quantile curve, specifically, the median ($\tau = 0.5$),  are reported in Table \ref{table:pairedMedian20} and, included in the Appendix, in Table \ref{table:pairedMedian40}, for 20\% and 40\%  of censored data, respectively. The proposed method is able to detect correctly the equality of the quantile curves, where, as before, the power of the test increases with increasing values of $\beta_{21} - \beta_{11}$ and $n$, or with decreasing percentages of censored data. The same holds when the covariance between the errors increases. The results for the different estimators and test statistics are comparable. As before, the test is conservative: the type I error is smaller than the theoretical value of $5\%$ in most cases. This is even more pronounced than for the comparison of distribution functions.  Despite this conservative behavior, the tests have good power properties.   

\begin{table}[H]
\centering
\begin{adjustbox}{totalheight=\textheight-3\baselineskip}
\small
\begin{tabular}{|c|c|c|ccc|ccc|ccc|ccc|} 
\cline{4-15}
\multicolumn{1}{c}{} & \multicolumn{1}{c}{} &       & Bonf  & $T_{L_2}$ & $T_{L_\infty}$ & Bonf  & $T_{L_2}$ & $T_{L_\infty}$ & Bonf  & $T_{L_2}$ & $T_{L_\infty}$ & Bonf  & $T_{L_2}$ & $T_{L_\infty}$  \\ 
\cline{4-15}
\multicolumn{1}{c}{} & \multicolumn{1}{c}{} &       & \multicolumn{3}{c|}{\cite{debacker2019}}     & \multicolumn{3}{c|}{\cite{peng2008survival}}     & \multicolumn{3}{c|}{\cite{portnoy2003censored}}       & \multicolumn{3}{c|}{\cite{wang2009locally}}       \\ 
\hline
Mod.                 & $n$                    & Diff. & \multicolumn{12}{c|}{$\eta = 0.2$}                                                                                                                 \\ 
\hline
\multirow{9}{*}{1}   & \multirow{3}{*}{100} & 0     & 0.028 & 0.014     & 0.020          & 0.040 & 0.036     & 0.030          & 0.048 & 0.032     & 0.034          & 0.054 & 0.014     & 0.023           \\
                     &                      & 0.2   & 0.060 & 0.022     & 0.028          & 0.076 & 0.038     & 0.052          & 0.044 & 0.022     & 0.026          & 0.100 & 0.043     & 0.048           \\
                     &                      & 0.4   & 0.230 & 0.178     & 0.184          & 0.171 & 0.114     & 0.094          & 0.162 & 0.218     & 0.168          & 0.179 & 0.173     & 0.179           \\ 
\cline{2-15}
                     & \multirow{3}{*}{200} & 0     & 0.034 & 0.024     & 0.016          & 0.030 & 0.020     & 0.024          & 0.046 & 0.030     & 0.030          & 0.029 & 0.016     & 0.016           \\
                     &                      & 0.2   & 0.116 & 0.084     & 0.082          & 0.134 & 0.098     & 0.104          & 0.152 & 0.138     & 0.124          & 0.092 & 0.104     & 0.086           \\
                     &                      & 0.4   & 0.454 & 0.476     & 0.448          & 0.470 & 0.512     & 0.486          & 0.458 & 0.540     & 0.482          & 0.495 & 0.503     & 0.461           \\ 
\cline{2-15}
                     & \multirow{3}{*}{400} & 0     & 0.048 & 0.034     & 0.034          & 0.046 & 0.046     & 0.040          & 0.036 & 0.016     & 0.032          & 0.052 & 0.032     & 0.026           \\
                     &                      & 0.2   & 0.296 & 0.248     & 0.236          & 0.266 & 0.268     & 0.234          & 0.266 & 0.282     & 0.266          & 0.242 & 0.246     & 0.212           \\
                     &                      & 0.4   & 0.862 & 0.864     & 0.858          & 0.794 & 0.838     & 0.792          & 0.826 & 0.896     & 0.864          & 0.816 & 0.872     & 0.790           \\ 
\hline
\multirow{9}{*}{2}   & \multirow{3}{*}{100} & 0     & 0.030 & 0.024     & 0.018          & 0.032 & 0.024     & 0.024          & 0.024 & 0.020     & 0.014          & 0.022 & 0.014     & 0.014           \\
                     &                      & 0.2   & 0.132 & 0.092     & 0.090          & 0.080 & 0.074     & 0.078          & 0.108 & 0.072     & 0.084          & 0.144 & 0.103     & 0.106           \\
                     &                      & 0.4   & 0.508 & 0.482     & 0.452          & 0.461 & 0.445     & 0.373          & 0.295 & 0.321     & 0.259          & 0.278 & 0.312     & 0.253           \\ 
\cline{2-15}
                     & \multirow{3}{*}{200} & 0     & 0.048 & 0.022     & 0.032          & 0.038 & 0.014     & 0.016          & 0.066 & 0.038     & 0.038          & 0.029 & 0.017     & 0.019           \\
                     &                      & 0.2   & 0.318 & 0.226     & 0.218          & 0.208 & 0.200     & 0.184          & 0.224 & 0.286     & 0.212          & 0.224 & 0.212     & 0.202           \\
                     &                      & 0.4   & 0.834 & 0.858     & 0.798          & 0.850 & 0.890     & 0.858          & 0.856 & 0.864     & 0.828          & 0.844 & 0.834     & 0.741           \\ 
\cline{2-15}
                     & \multirow{3}{*}{400} & 0     & 0.046 & 0.020     & 0.030          & 0.060 & 0.044     & 0.034          & 0.058 & 0.038     & 0.050          & 0.064 & 0.032     & 0.042           \\
                     &                      & 0.2   & 0.528 & 0.466     & 0.488          & 0.508 & 0.554     & 0.514          & 0.604 & 0.566     & 0.536          & 0.472 & 0.518     & 0.474           \\
                     &                      & 0.4   & 0.988 & 0.996     & 0.988          & 0.996 & 0.998     & 0.998          & 0.992 & 0.996     & 0.994          & 0.996 & 0.992     & 0.994           \\ 
\hline
\multirow{9}{*}{3}   & \multirow{3}{*}{100} & 0     & 0.036 & 0.022     & 0.028          & 0.032 & 0.010     & 0.012          & 0.030 & 0.016     & 0.016          & 0.036 & 0.024     & 0.028           \\
                     &                      & 0.2   & 0.192 & 0.110     & 0.120          & 0.210 & 0.176     & 0.170          & 0.198 & 0.140     & 0.152          & 0.211 & 0.188     & 0.176           \\
                     &                      & 0.4   & 0.830 & 0.814     & 0.792          & 0.748 & 0.806     & 0.756          & 0.776 & 0.772     & 0.742          & 0.723 & 0.762     & 0.739           \\ 
\cline{2-15}
                     & \multirow{3}{*}{200} & 0     & 0.048 & 0.040     & 0.042          & 0.036 & 0.028     & 0.024          & 0.058 & 0.036     & 0.052          & 0.040 & 0.026     & 0.024           \\
                     &                      & 0.2   & 0.354 & 0.460     & 0.398          & 0.542 & 0.504     & 0.468          & 0.440 & 0.510     & 0.478          & 0.428 & 0.392     & 0.432           \\
                     &                      & 0.4   & 0.992 & 0.992     & 0.992          & 0.988 & 0.994     & 0.988          & 0.988 & 0.996     & 0.986          & 0.984 & 0.990     & 0.982           \\ 
\cline{2-15}
                     & \multirow{3}{*}{400} & 0     & 0.050 & 0.016     & 0.028          & 0.060 & 0.044     & 0.048          & 0.040 & 0.036     & 0.044          & 0.036 & 0.034     & 0.038           \\
                     &                      & 0.2   & 0.836 & 0.860     & 0.820          & 0.822 & 0.874     & 0.838          & 0.838 & 0.856     & 0.820          & 0.784 & 0.854     & 0.844           \\
                     &                      & 0.4   & 1.000 & 1.000     & 1.000          & 1.000 & 1.000     & 1.000          & 1.000 & 1.000     & 1.000          & 1.000 & 1.000     & 1.000           \\ 
\hline
\multicolumn{3}{|c|}{}       & \multicolumn{12}{c|}{$\eta = 0.4$}                                                                                                                 \\ 
\hline
\multirow{9}{*}{1}   & \multirow{3}{*}{100} & 0     & 0.046 & 0.010     & 0.014          & 0.042 & 0.020     & 0.026          & 0.016 & 0.002     & 0.010          & 0.018 & 0.005     & 0.009           \\
                     &                      & 0.2   & 0.082 & 0.056     & 0.046          & 0.072 & 0.060     & 0.054          & 0.056 & 0.054     & 0.050          & 0.073 & 0.068     & 0.073           \\
                     &                      & 0.4   & 0.190 & 0.206     & 0.176          & 0.242 & 0.228     & 0.186          & 0.293 & 0.315     & 0.267          & 0.325 & 0.247     & 0.226           \\ 
\cline{2-15}
                     & \multirow{3}{*}{200} & 0     & 0.028 & 0.020     & 0.020          & 0.026 & 0.010     & 0.014          & 0.036 & 0.022     & 0.022          & 0.053 & 0.024     & 0.032           \\
                     &                      & 0.2   & 0.212 & 0.184     & 0.174          & 0.186 & 0.146     & 0.148          & 0.170 & 0.118     & 0.156          & 0.162 & 0.133     & 0.131           \\
                     &                      & 0.4   & 0.740 & 0.670     & 0.648          & 0.676 & 0.728     & 0.670          & 0.678 & 0.694     & 0.666          & 0.563 & 0.569     & 0.523           \\ 
\cline{2-15}
                     & \multirow{3}{*}{400} & 0     & 0.072 & 0.032     & 0.040          & 0.050 & 0.036     & 0.038          & 0.036 & 0.024     & 0.018          & 0.026 & 0.026     & 0.018           \\
                     &                      & 0.2   & 0.428 & 0.388     & 0.372          & 0.348 & 0.320     & 0.286          & 0.374 & 0.410     & 0.376          & 0.342 & 0.346     & 0.344           \\
                     &                      & 0.4   & 0.940 & 0.946     & 0.920          & 0.976 & 0.984     & 0.962          & 0.968 & 0.976     & 0.968          & 0.954 & 0.976     & 0.962           \\ 
\hline
\multirow{9}{*}{2}   & \multirow{3}{*}{100} & 0     & 0.038 & 0.004     & 0.014          & 0.022 & 0.012     & 0.008          & 0.024 & 0.010     & 0.012          & 0.022 & 0.007     & 0.005           \\
                     &                      & 0.2   & 0.110 & 0.106     & 0.068          & 0.092 & 0.076     & 0.068          & 0.082 & 0.064     & 0.062          & 0.120 & 0.074     & 0.066           \\
                     &                      & 0.4   & 0.518 & 0.484     & 0.434          & 0.628 & 0.604     & 0.519          & 0.490 & 0.566     & 0.474          & 0.473 & 0.482     & 0.430           \\ 
\cline{2-15}
                     & \multirow{3}{*}{200} & 0     & 0.046 & 0.022     & 0.022          & 0.038 & 0.026     & 0.028          & 0.028 & 0.012     & 0.020          & 0.047 & 0.025     & 0.023           \\
                     &                      & 0.2   & 0.306 & 0.276     & 0.218          & 0.320 & 0.274     & 0.276          & 0.418 & 0.394     & 0.352          & 0.352 & 0.305     & 0.293           \\
                     &                      & 0.4   & 0.940 & 0.944     & 0.922          & 0.958 & 0.966     & 0.954          & 0.948 & 0.964     & 0.948          & 0.909 & 0.938     & 0.922           \\ 
\cline{2-15}
                     & \multirow{3}{*}{400} & 0     & 0.046 & 0.030     & 0.026          & 0.044 & 0.028     & 0.032          & 0.042 & 0.026     & 0.034          & 0.030 & 0.016     & 0.016           \\
                     &                      & 0.2   & 0.720 & 0.710     & 0.690          & 0.722 & 0.706     & 0.684          & 0.770 & 0.798     & 0.756          & 0.734 & 0.720     & 0.692           \\
                     &                      & 0.4   & 1.000 & 1.000     & 1.000          & 1.000 & 1.000     & 1.000          & 1.000 & 1.000     & 1.000          & 1.000 & 1.000     & 1.000           \\ 
\hline
\multirow{9}{*}{3}   & \multirow{3}{*}{100} & 0     & 0.036 & 0.008     & 0.016          & 0.016 & 0.004     & 0.014          & 0.024 & 0.010     & 0.010          & 0.024 & 0.006     & 0.008           \\
                     &                      & 0.2   & 0.354 & 0.280     & 0.264          & 0.318 & 0.288     & 0.270          & 0.234 & 0.218     & 0.216          & 0.310 & 0.239     & 0.237           \\
                     &                      & 0.4   & 0.886 & 0.912     & 0.876          & 0.928 & 0.912     & 0.872          & 0.922 & 0.942     & 0.916          & 0.915 & 0.943     & 0.929           \\ 
\cline{2-15}
                     & \multirow{3}{*}{200} & 0     & 0.056 & 0.040     & 0.042          & 0.028 & 0.012     & 0.014          & 0.042 & 0.042     & 0.034          & 0.026 & 0.012     & 0.018           \\
                     &                      & 0.2   & 0.586 & 0.650     & 0.572          & 0.730 & 0.720     & 0.632          & 0.592 & 0.632     & 0.596          & 0.554 & 0.502     & 0.488           \\
                     &                      & 0.4   & 0.998 & 1.000     & 0.998          & 1.000 & 1.000     & 1.000          & 0.998 & 1.000     & 0.996          & 1.000 & 1.000     & 0.998           \\ 
\cline{2-15}
                     & \multirow{3}{*}{400} & 0     & 0.034 & 0.020     & 0.034          & 0.052 & 0.032     & 0.034          & 0.044 & 0.032     & 0.032          & 0.046 & 0.020     & 0.022           \\
                     &                      & 0.2   & 0.952 & 0.960     & 0.956          & 0.904 & 0.944     & 0.914          & 0.960 & 0.966     & 0.940          & 0.954 & 0.960     & 0.954           \\
                     &                      & 0.4   & 1.000 & 1.000     & 1.000          & 1.000 & 1.000     & 1.000          & 1.000 & 1.000     & 1.000          & 1.000 & 1.000     & 1.000           \\
\hline
\end{tabular}
\end{adjustbox}
\caption{\label{table:pairedDistribution20}Simulation results for the comparison of quantile curves (range of $\tau$-values) for two paired samples for percentage of censoring equal to 20\%.  The sample size is $n=100,200,400$, the significance level is $\alpha=0.05$ and Diff $=\beta_{21}-\beta_{11}$, so the null hypothesis is satisfied for Diff $=0$. The covariance between the errors is indicated by $\eta$, see \eqref{eq:dependent}.  The abbreviations  Bonf, $T_{L_2}$ and $T_{L_\infty}$ stand for the test statistics defined in (\ref{Bonf}), (\ref{L2}) and (\ref{Linfty}), respectively. }
\end{table}

\begin{table}[H]
\centering
\begin{adjustbox}{totalheight=\textheight-3\baselineskip}
\small
\begin{tabular}{|c|c|c|ccc|ccc|ccc|ccc|} 
\cline{4-15}
\multicolumn{1}{c}{} & \multicolumn{1}{c}{} &       & Bonf  & $T_{L_2}$ & $T_{L_\infty}$ & Bonf  & $T_{L_2}$ & $T_{L_\infty}$ & Bonf  & $T_{L_2}$ & $T_{L_\infty}$ & Bonf  & $T_{L_2}$ & $T_{L_\infty}$  \\ 
\cline{4-15}
\multicolumn{1}{c}{} & \multicolumn{1}{c}{} &       & \multicolumn{3}{c|}{\cite{debacker2019}}     & \multicolumn{3}{c|}{\cite{peng2008survival}}     & \multicolumn{3}{c|}{\cite{portnoy2003censored}}       & \multicolumn{3}{c|}{\cite{wang2009locally}}       \\ 
\hline
Mod.                 & $n$                    & Diff. & \multicolumn{12}{c|}{$\eta = 0.2$}                                                                                                                 \\ 
\hline
\multirow{9}{*}{1}   & \multirow{3}{*}{100} & 0     & 0.022 & 0.024     & 0.022          & 0.010 & 0.004     & 0.010          & 0.020 & 0.016     & 0.020          & 0.020 & 0.012     & 0.016           \\
                     &                      & 0.2   & 0.018 & 0.018     & 0.014          & 0.024 & 0.028     & 0.028          & 0.008 & 0.012     & 0.016          & 0.034 & 0.030     & 0.020           \\
                     &                      & 0.4   & 0.136 & 0.118     & 0.136          & 0.082 & 0.064     & 0.074          & 0.112 & 0.106     & 0.128          & 0.088 & 0.078     & 0.056           \\ 
\cline{2-15}
                     & \multirow{3}{*}{200} & 0     & 0.010 & 0.012     & 0.008          & 0.014 & 0.012     & 0.014          & 0.020 & 0.028     & 0.020          & 0.016 & 0.016     & 0.014           \\
                     &                      & 0.2   & 0.058 & 0.056     & 0.062          & 0.074 & 0.090     & 0.082          & 0.086 & 0.084     & 0.086          & 0.038 & 0.050     & 0.040           \\
                     &                      & 0.4   & 0.276 & 0.298     & 0.228          & 0.200 & 0.248     & 0.236          & 0.302 & 0.310     & 0.290          & 0.260 & 0.392     & 0.288           \\ 
\cline{2-15}
                     & \multirow{3}{*}{400} & 0     & 0.040 & 0.034     & 0.042          & 0.024 & 0.036     & 0.028          & 0.020 & 0.018     & 0.014          & 0.014 & 0.014     & 0.014           \\
                     &                      & 0.2   & 0.158 & 0.168     & 0.124          & 0.140 & 0.148     & 0.122          & 0.138 & 0.164     & 0.152          & 0.154 & 0.114     & 0.158           \\
                     &                      & 0.4   & 0.620 & 0.710     & 0.640          & 0.562 & 0.636     & 0.554          & 0.658 & 0.682     & 0.652          & 0.508 & 0.616     & 0.510           \\ 
\hline
\multirow{9}{*}{2}   & \multirow{3}{*}{100} & 0     & 0.016 & 0.022     & 0.010          & 0.020 & 0.006     & 0.020          & 0.014 & 0.014     & 0.008          & 0.024 & 0.012     & 0.024           \\
                     &                      & 0.2   & 0.022 & 0.028     & 0.020          & 0.024 & 0.018     & 0.024          & 0.036 & 0.032     & 0.030          & 0.065 & 0.062     & 0.056           \\
                     &                      & 0.4   & 0.172 & 0.196     & 0.152          & 0.156 & 0.182     & 0.188          & 0.270 & 0.282     & 0.276          & 0.181 & 0.301     & 0.161           \\ 
\cline{2-15}
                     & \multirow{3}{*}{200} & 0     & 0.032 & 0.024     & 0.030          & 0.022 & 0.020     & 0.022          & 0.018 & 0.024     & 0.018          & 0.012 & 0.014     & 0.012           \\
                     &                      & 0.2   & 0.062 & 0.068     & 0.060          & 0.110 & 0.134     & 0.110          & 0.140 & 0.168     & 0.152          & 0.110 & 0.094     & 0.112           \\
                     &                      & 0.4   & 0.572 & 0.652     & 0.560          & 0.606 & 0.678     & 0.598          & 0.616 & 0.654     & 0.616          & 0.522 & 0.626     & 0.540           \\ 
\cline{2-15}
                     & \multirow{3}{*}{400} & 0     & 0.022 & 0.020     & 0.022          & 0.042 & 0.042     & 0.038          & 0.024 & 0.020     & 0.022          & 0.028 & 0.036     & 0.030           \\
                     &                      & 0.2   & 0.346 & 0.380     & 0.272          & 0.392 & 0.420     & 0.352          & 0.278 & 0.282     & 0.260          & 0.268 & 0.282     & 0.208           \\
                     &                      & 0.4   & 0.948 & 0.962     & 0.940          & 0.950 & 0.986     & 0.944          & 0.958 & 0.976     & 0.954          & 0.884 & 0.944     & 0.896           \\ 
\hline
\multirow{9}{*}{3}   & \multirow{3}{*}{100} & 0     & 0.020 & 0.012     & 0.018          & 0.006 & 0.000     & 0.000          & 0.010 & 0.008     & 0.016          & 0.014 & 0.010     & 0.004           \\
                     &                      & 0.2   & 0.072 & 0.082     & 0.080          & 0.170 & 0.142     & 0.156          & 0.120 & 0.080     & 0.120          & 0.100 & 0.096     & 0.086           \\
                     &                      & 0.4   & 0.422 & 0.552     & 0.470          & 0.516 & 0.562     & 0.534          & 0.500 & 0.584     & 0.484          & 0.456 & 0.554     & 0.444           \\ 
\cline{2-15}
                     & \multirow{3}{*}{200} & 0     & 0.038 & 0.024     & 0.040          & 0.014 & 0.020     & 0.018          & 0.020 & 0.020     & 0.018          & 0.022 & 0.022     & 0.024           \\
                     &                      & 0.2   & 0.148 & 0.204     & 0.146          & 0.224 & 0.252     & 0.238          & 0.228 & 0.262     & 0.224          & 0.294 & 0.346     & 0.298           \\
                     &                      & 0.4   & 0.838 & 0.906     & 0.860          & 0.896 & 0.944     & 0.918          & 0.890 & 0.972     & 0.910          & 0.870 & 0.948     & 0.880           \\ 
\cline{2-15}
                     & \multirow{3}{*}{400} & 0     & 0.030 & 0.018     & 0.028          & 0.018 & 0.036     & 0.022          & 0.012 & 0.014     & 0.010          & 0.032 & 0.028     & 0.032           \\
                     &                      & 0.2   & 0.546 & 0.662     & 0.564          & 0.580 & 0.642     & 0.584          & 0.538 & 0.664     & 0.566          & 0.592 & 0.644     & 0.560           \\
                     &                      & 0.4   & 1.000 & 1.000     & 1.000          & 1.000 & 1.000     & 1.000          & 1.000 & 1.000     & 1.000          & 1.000 & 1.000     & 1.000           \\ 
\hline
\multicolumn{3}{|c|}{}      & \multicolumn{12}{c|}{$\eta = 0.4$}                                                                                                                 \\ 
\hline
\multirow{9}{*}{1}   & \multirow{3}{*}{100} & 0     & 0.010 & 0.012     & 0.012          & 0.006 & 0.006     & 0.004          & 0.008 & 0.004     & 0.008          & 0.014 & 0.014     & 0.014           \\
                     &                      & 0.2   & 0.044 & 0.030     & 0.036          & 0.024 & 0.016     & 0.024          & 0.014 & 0.022     & 0.014          & 0.022 & 0.030     & 0.032           \\
                     &                      & 0.4   & 0.050 & 0.060     & 0.062          & 0.064 & 0.082     & 0.084          & 0.070 & 0.056     & 0.076          & 0.186 & 0.184     & 0.188           \\ 
\cline{2-15}
                     & \multirow{3}{*}{200} & 0     & 0.010 & 0.018     & 0.010          & 0.022 & 0.012     & 0.016          & 0.026 & 0.016     & 0.016          & 0.026 & 0.014     & 0.016           \\
                     &                      & 0.2   & 0.040 & 0.088     & 0.046          & 0.076 & 0.056     & 0.072          & 0.062 & 0.078     & 0.062          & 0.074 & 0.074     & 0.072           \\
                     &                      & 0.4   & 0.350 & 0.416     & 0.342          & 0.252 & 0.274     & 0.220          & 0.260 & 0.406     & 0.306          & 0.328 & 0.340     & 0.334           \\ 
\cline{2-15}
                     & \multirow{3}{*}{400} & 0     & 0.026 & 0.018     & 0.022          & 0.026 & 0.014     & 0.028          & 0.018 & 0.010     & 0.014          & 0.006 & 0.012     & 0.006           \\
                     &                      & 0.2   & 0.242 & 0.242     & 0.240          & 0.146 & 0.200     & 0.164          & 0.218 & 0.282     & 0.214          & 0.270 & 0.212     & 0.252           \\
                     &                      & 0.4   & 0.716 & 0.720     & 0.712          & 0.760 & 0.818     & 0.792          & 0.722 & 0.804     & 0.722          & 0.696 & 0.796     & 0.732           \\ 
\hline
\multirow{9}{*}{2}   & \multirow{3}{*}{100} & 0     & 0.012 & 0.008     & 0.014          & 0.012 & 0.010     & 0.006          & 0.006 & 0.014     & 0.012          & 0.012 & 0.008     & 0.010           \\
                     &                      & 0.2   & 0.044 & 0.022     & 0.034          & 0.042 & 0.042     & 0.038          & 0.026 & 0.040     & 0.028          & 0.026 & 0.020     & 0.022           \\
                     &                      & 0.4   & 0.174 & 0.236     & 0.224          & 0.152 & 0.210     & 0.182          & 0.188 & 0.186     & 0.172          & 0.205 & 0.313     & 0.215           \\ 
\cline{2-15}
                     & \multirow{3}{*}{200} & 0     & 0.030 & 0.022     & 0.032          & 0.010 & 0.010     & 0.010          & 0.014 & 0.008     & 0.018          & 0.028 & 0.010     & 0.028           \\
                     &                      & 0.2   & 0.190 & 0.216     & 0.190          & 0.092 & 0.124     & 0.094          & 0.166 & 0.164     & 0.174          & 0.128 & 0.150     & 0.128           \\
                     &                      & 0.4   & 0.690 & 0.808     & 0.720          & 0.674 & 0.752     & 0.648          & 0.802 & 0.840     & 0.778          & 0.696 & 0.764     & 0.724           \\ 
\cline{2-15}
                     & \multirow{3}{*}{400} & 0     & 0.016 & 0.006     & 0.012          & 0.028 & 0.016     & 0.026          & 0.002 & 0.008     & 0.006          & 0.018 & 0.010     & 0.016           \\
                     &                      & 0.2   & 0.464 & 0.534     & 0.488          & 0.422 & 0.434     & 0.432          & 0.430 & 0.480     & 0.430          & 0.394 & 0.420     & 0.400           \\
                     &                      & 0.4   & 0.968 & 0.982     & 0.968          & 0.988 & 0.998     & 0.988          & 0.978 & 0.990     & 0.972          & 0.950 & 0.992     & 0.954           \\ 
\hline
\multirow{9}{*}{3}   & \multirow{3}{*}{100} & 0     & 0.018 & 0.014     & 0.014          & 0.006 & 0.004     & 0.006          & 0.012 & 0.010     & 0.014          & 0.016 & 0.010     & 0.016           \\
                     &                      & 0.2   & 0.094 & 0.122     & 0.080          & 0.116 & 0.154     & 0.118          & 0.098 & 0.110     & 0.106          & 0.084 & 0.070     & 0.092           \\
                     &                      & 0.4   & 0.614 & 0.652     & 0.620          & 0.490 & 0.616     & 0.514          & 0.618 & 0.678     & 0.598          & 0.466 & 0.536     & 0.406           \\ 
\cline{2-15}
                     & \multirow{3}{*}{200} & 0     & 0.012 & 0.028     & 0.012          & 0.032 & 0.020     & 0.022          & 0.038 & 0.030     & 0.036          & 0.012 & 0.010     & 0.012           \\
                     &                      & 0.2   & 0.334 & 0.454     & 0.350          & 0.324 & 0.406     & 0.336          & 0.318 & 0.374     & 0.354          & 0.306 & 0.396     & 0.302           \\
                     &                      & 0.4   & 0.952 & 0.984     & 0.960          & 0.960 & 0.994     & 0.974          & 0.958 & 0.990     & 0.974          & 0.938 & 0.970     & 0.938           \\ 
\cline{2-15}
                     & \multirow{3}{*}{400} & 0     & 0.018 & 0.016     & 0.018          & 0.032 & 0.036     & 0.032          & 0.030 & 0.010     & 0.030          & 0.018 & 0.014     & 0.014           \\
                     &                      & 0.2   & 0.682 & 0.782     & 0.692          & 0.700 & 0.792     & 0.688          & 0.688 & 0.798     & 0.682          & 0.690 & 0.814     & 0.714           \\
                     &                      & 0.4   & 1.000 & 1.000     & 1.000          & 1.000 & 1.000     & 1.000          & 1.000 & 1.000     & 1.000          & 1.000 & 1.000     & 1.000           \\
\hline
\end{tabular}
\end{adjustbox}
\caption{\label{table:pairedMedian20}Simulation results for the comparison of the median ($\tau=0.5$) for two paired samples subject to $20\%$ censoring. The sample size is $n=100,200,400$, the significance level is $\alpha=0.05$ and Diff $=\beta_{21}-\beta_{11}$, so the null hypothesis is satisfied for Diff $=0$. The covariance between the errors is indicated by $\eta$, see \eqref{eq:dependent}. The abbreviations Bonf, $T_{L_2}$ and $T_{L_\infty}$ stand for the test statistics defined in (\ref{Bonf}), (\ref{L2}) and (\ref{Linfty}), respectively. }
\end{table}

\section{Data analysis} \label{sec7}

In this section we present an application of the proposed method using a dataset regarding diabetic retinopathy, see \cite{survival-package}. Studying this dataset is particularly interesting, since it will require the application of our testing procedure both in the cases of independent and paired samples. Moreover, it corresponds to a well-known case studied in the literature, which makes it possible to compare our results with previous results. Specifically, we will refer to \cite{huster1989modelling}. 

Diabetic retinopathy is a complication associated with diabetes mellitus consisting of abnormalities in the microvasculature within the retina of the eye. The Diabetic Retinopathy Study (DRS) begun in 1971 to study the effectiveness of laser photocoagulation in delaying the onset of blindness in patients with diabetic retinopathy.  Patients with diabetic retinopathy in both eyes and visual acuity of 20/100 were eligible for the study. One eye of each patient was randomly selected for treatment and the other eye was observed without treatment. For each eye, the event of interest is the time from initiation of treatment to the time when visual acuity dropped below 5/200 two visits in a row. Thus there is a built-in lag time of approximately 6 months (visits were every 3 months). Survival times in this dataset are therefore the actual time to blindness in months, minus the minimum possible time to event (6.5 months). Censoring is caused by death, dropout, or end of the study. The dataset includes $n = 197$ high-risk patients, as defined by DRS criteria.

It is important to notice that diabetes can be classified into two general groups by the age at onset: juvenile and adult diabetes. The two diseases have, in general, very different courses. The primary question of the DRS study was to assess the effectiveness of the laser photocoagulation treatment. Moreover, we would like to know whether the effect of the treatment depends on the type of diabetes. Lastly, since for the treated eyes \textit{xenon} and \textit{argon} lasers were randomly used, we would also like to understand which laser therapy is most effective. 

We now use our method for finding answers to these questions. 
In order to assess the effectiveness of the laser treatment, we apply our method in the case of paired samples. We compare the distributions of two samples which include, respectively, the data regarding the treated eyes and the control eyes (not treated eyes) of the $197$ patients. Due to the specifications of the experiment, the censoring times for the two eyes are equal. The samples contain $60\%$ of censored data, which limits the range of quantile levels we can use in our analysis.  Therefore, we compare the quantiles for $\tau \in [0.1, 0.3]$, with step size equal to $0.01$. For the comparison, we chose the estimator of \cite{peng2008survival}, and we use the same standardization of the test statistics as in the simulation study.

\begin{figure}[H]
\centering     
\subfigure[]{\label{fig:a}\includegraphics[width=80mm]{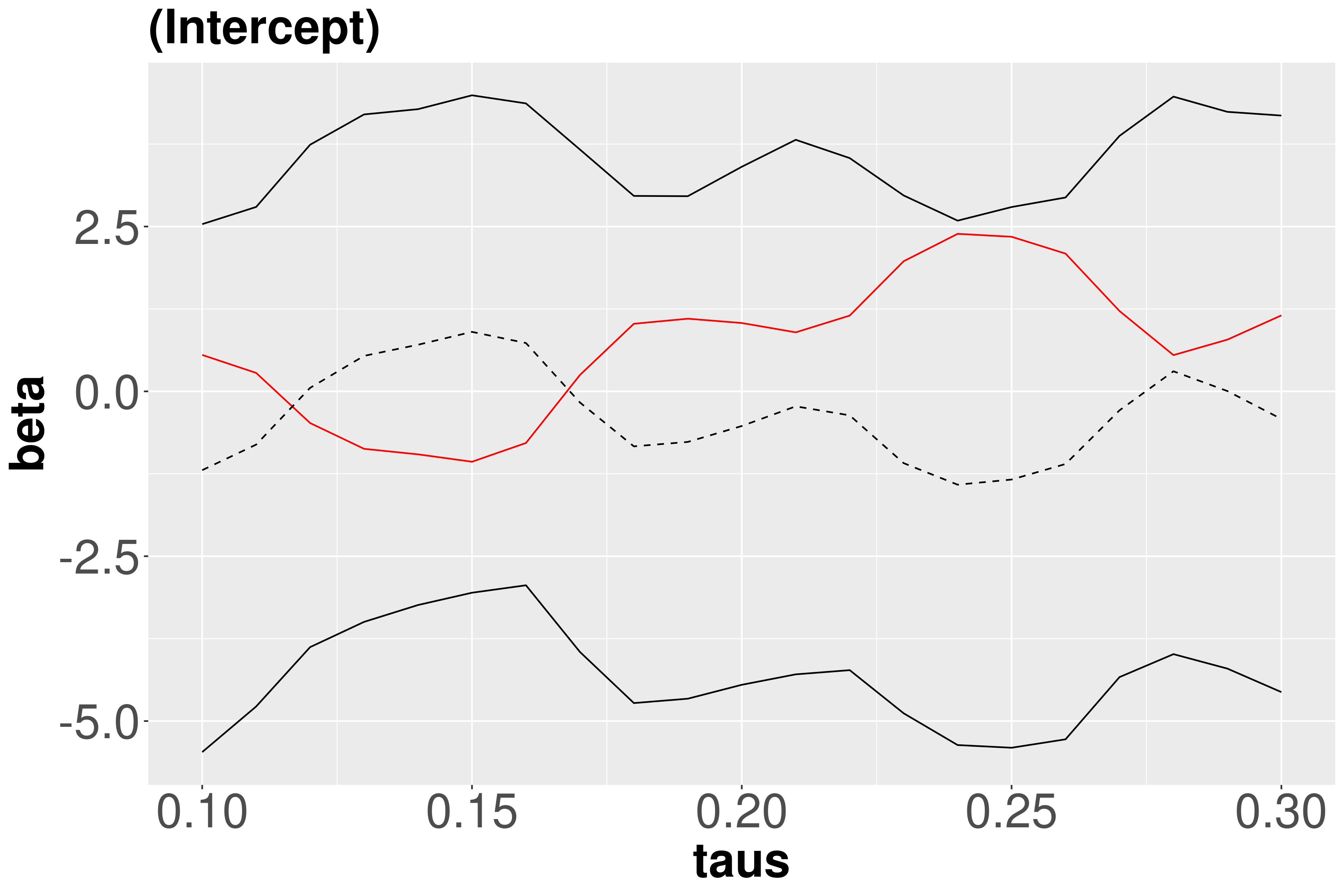}}
\subfigure[]{\label{fig:b}\includegraphics[width=80mm]{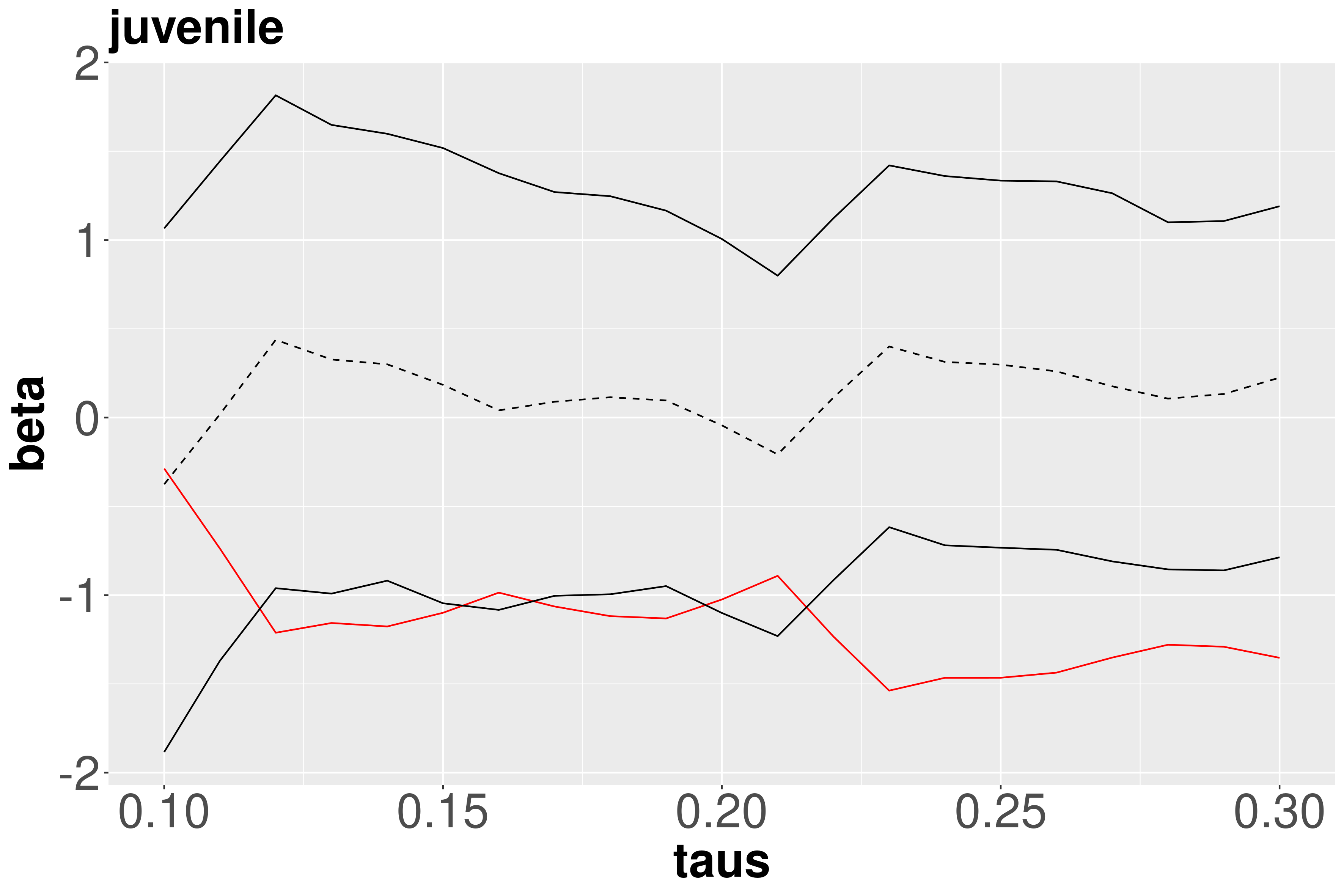}}
\subfigure[]{\includegraphics[width=80mm]{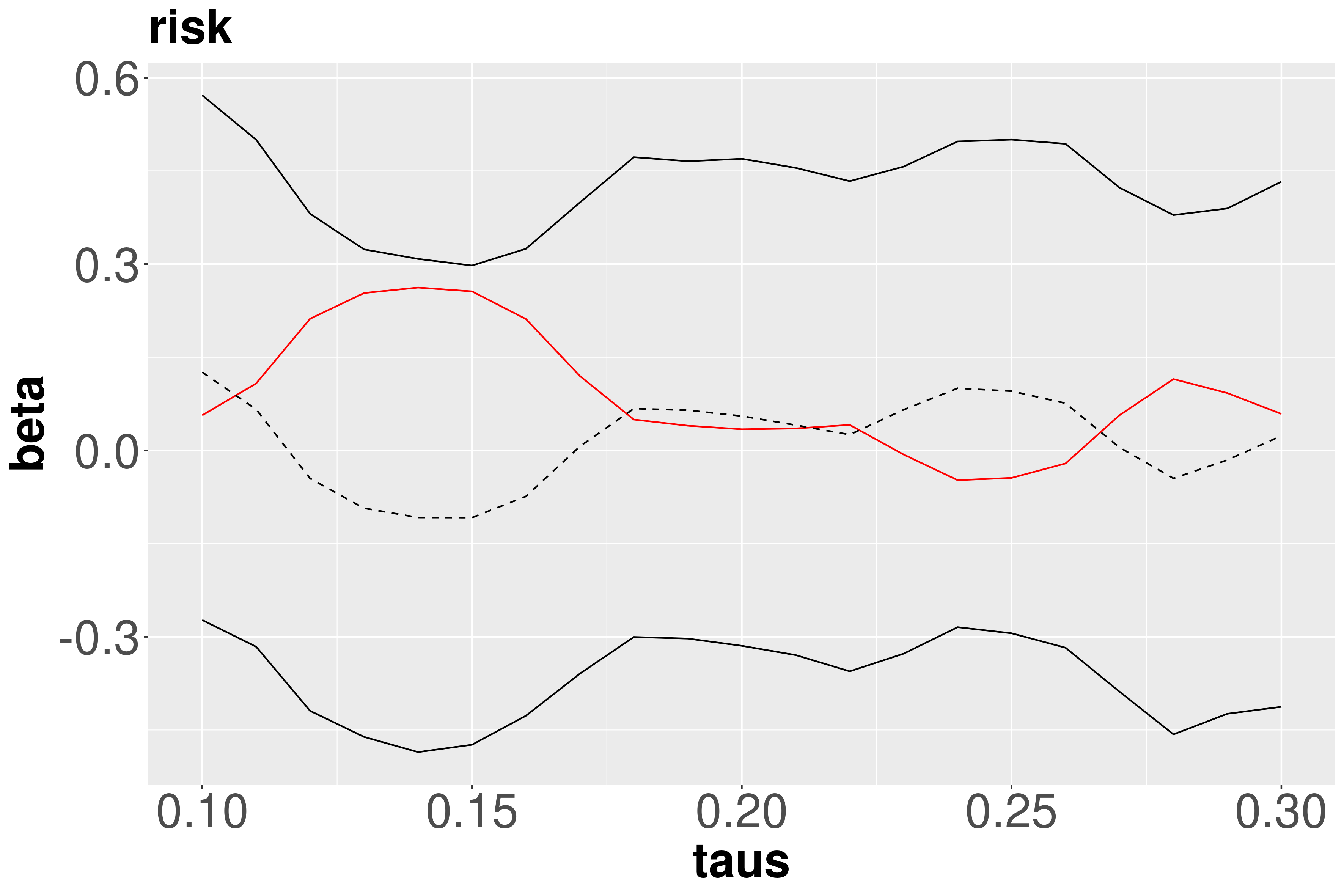}}
\caption{\label{fig:cov3}  The process $\{\hat\beta_{\text{treated}} - \hat\beta_{\text{control}}\}(\tau)$ for each coefficient (in red), together with $2.5\%$ and $97.5\%$ percentiles of the distribution of the bootstrap process 
$\{( \hat\beta_{\text{treated}}^* - \hat\beta_{\text{control}}^*) - (\hat\beta_{\text{treated}} - \hat\beta_{\text{control}}) \}(\tau)$ (black solid lines) and the mean bootstrap process (dotted black line). Subfigures (a), (b) and (c) show the estimated intercept and the estimated coefficients of \textit{juvenile} and \textit{risk} respectively, for a range of $\tau$-values. }
\end{figure}

For inference purposes, we use the same percentile bootstrap procedure as in the simulation study, where the number of bootstrap resamples is $N = 10000$. We use the re-weighting bootstrap method discussed in Section \ref{sec5}.  We include the \textit{juvenile} covariate, which equals 1 for juveline diabetes and 0 otherwise,  and the \textit{risk} covariate, which indicates the risk level of a patient as specified by the criteria of RDS and takes discrete values from 6 to 12. We present the results both graphically and numerically, respectively in Figure \ref{fig:cov3} and Table \ref{table:covcoeff3}.

\begin{table}[H]
\centering
\begin{tabular}{|c|ccccc|}
\hline
            & \multicolumn{1}{c|}{$\tau= 0.10$} & \multicolumn{1}{c|}{$\tau= 0.15$} & \multicolumn{1}{c|}{$\tau= 0.20$} & \multicolumn{1}{c|}{$\tau= 0.25$} & $\tau= 0.30$ \\ \hline
            & \multicolumn{5}{c|}{$\hat\beta_{\text{treated}}$}                                                                                                            \\ \hline
intercept & \multicolumn{1}{c|}{4.64}      & \multicolumn{1}{c|}{3.78}      & \multicolumn{1}{c|}{5.71}      & \multicolumn{1}{c|}{7.36}      & 6.61      \\ \hline
juvenile    & \multicolumn{1}{c|}{-0.37}     & \multicolumn{1}{c|}{-0.67}     & \multicolumn{1}{c|}{-0.81}     & \multicolumn{1}{c|}{-1.27}     & -0.85     \\ \hline
risk        & \multicolumn{1}{c|}{-0.21}     & \multicolumn{1}{c|}{-0.05}     & \multicolumn{1}{c|}{-0.21}     & \multicolumn{1}{c|}{-0.29}     & -0.22     \\ \hline
            & \multicolumn{5}{c|}{$\hat\beta_{\text{control}}$}                                                                                                            \\ \hline
intercept & \multicolumn{1}{c|}{4.09}      & \multicolumn{1}{c|}{4.85}      & \multicolumn{1}{c|}{4.68}      & \multicolumn{1}{c|}{5.01}      & 5.45      \\ \hline
juvenile    & \multicolumn{1}{c|}{-0.08}     & \multicolumn{1}{c|}{0.43}      & \multicolumn{1}{c|}{0.22}      & \multicolumn{1}{c|}{0.20}      & 0.50      \\ \hline
risk        & \multicolumn{1}{c|}{-0.27}     & \multicolumn{1}{c|}{-0.30}     & \multicolumn{1}{c|}{-0.24}     & \multicolumn{1}{c|}{-0.25}     & -0.28     \\ \hline
            & \multicolumn{5}{c|}{$\hat\beta_{\text{treated}}-\hat\beta_{\text{control}}$}                                                                                                \\ \hline
intercept & \multicolumn{1}{c|}{0.55}      & \multicolumn{1}{c|}{-1.07}     & \multicolumn{1}{c|}{1.04}      & \multicolumn{1}{c|}{2.34}      & 1.15      \\ \hline
juvenile    & \multicolumn{1}{c|}{-0.29}     & \multicolumn{1}{c|}{-1.10}     & \multicolumn{1}{c|}{-1.02}     & \multicolumn{1}{c|}{-1.47}     & -1.35     \\ \hline
risk        & \multicolumn{1}{c|}{0.06}      & \multicolumn{1}{c|}{0.26}      & \multicolumn{1}{c|}{0.03}      & \multicolumn{1}{c|}{-0.04}     & 0.06      \\ \hline
\end{tabular}
\caption{\label{table:covcoeff3} Values of the components intercept, \textit{juvenile} and \textit{risk} of the processes $\hat\beta_{\text{treated}}(\tau)$, $\hat\beta_{\text{control}}(\tau)$ and their difference, for $\tau\in\{0.10,0.15,...,0.30\}$.}
\end{table}

We can assess the effectiveness of the laser treatment by testing the hypothesis $H_0: \beta_{\text{treated}}(\tau) = \beta_{\text{control}}(\tau)$ for $\tau \in A = [0.1,0.3]$, using the test statistics presented in Section   \ref{sec3}.  For any of the test statistics $T_{L_2}, T_{L_\infty}$ and Bonferroni, the respective p-value is less than 0.05. Therefore, we can conclude (at level of confidence 0.95) that there is statistical evidence of difference between the treatment and control groups. We can also conclude that the quantiles for the treated eyes are larger than the quantiles for the control eyes. Indeed, we can observe that for each fixed covariate vector $Z= (Z_{\text{Intercept}}, Z_{\text{juvenile}}, Z_{\text{risk}})\in \{1\}\times\{0,1\}\times \{6,...,12\} $, we have  $\hat\beta_{\text{treated}}(\tau)Z \ge \hat\beta_{\text{control}}(\tau)Z$. 

Using the Delta method we can also verify whether there exists a component $k=1,2,3$ for which the null hypothesis $H_0: \beta_{\text{treated,k}}(\tau) =\beta_{\text{control,k}}(\tau)$ for $\tau\in A$ is rejected  (at level 0.05).
Such null hypothesis is rejected only for the \textit{juvenile} component (with p-value equal to 0.02), and so we can conclude that there is a significant difference in the effectiveness of the treatment for young and adult types of diabetes. Since $\hat{\beta}_{\text{treated,juvenile}} (\tau)< 0$, we can also note that the treatment is more effective for the adult type of diabetes. All our conclusions agree with \cite{huster1989modelling}. 

\begin{figure}[H]
\centering     
\subfigure[]{\label{fig:a}\includegraphics[width=80mm]{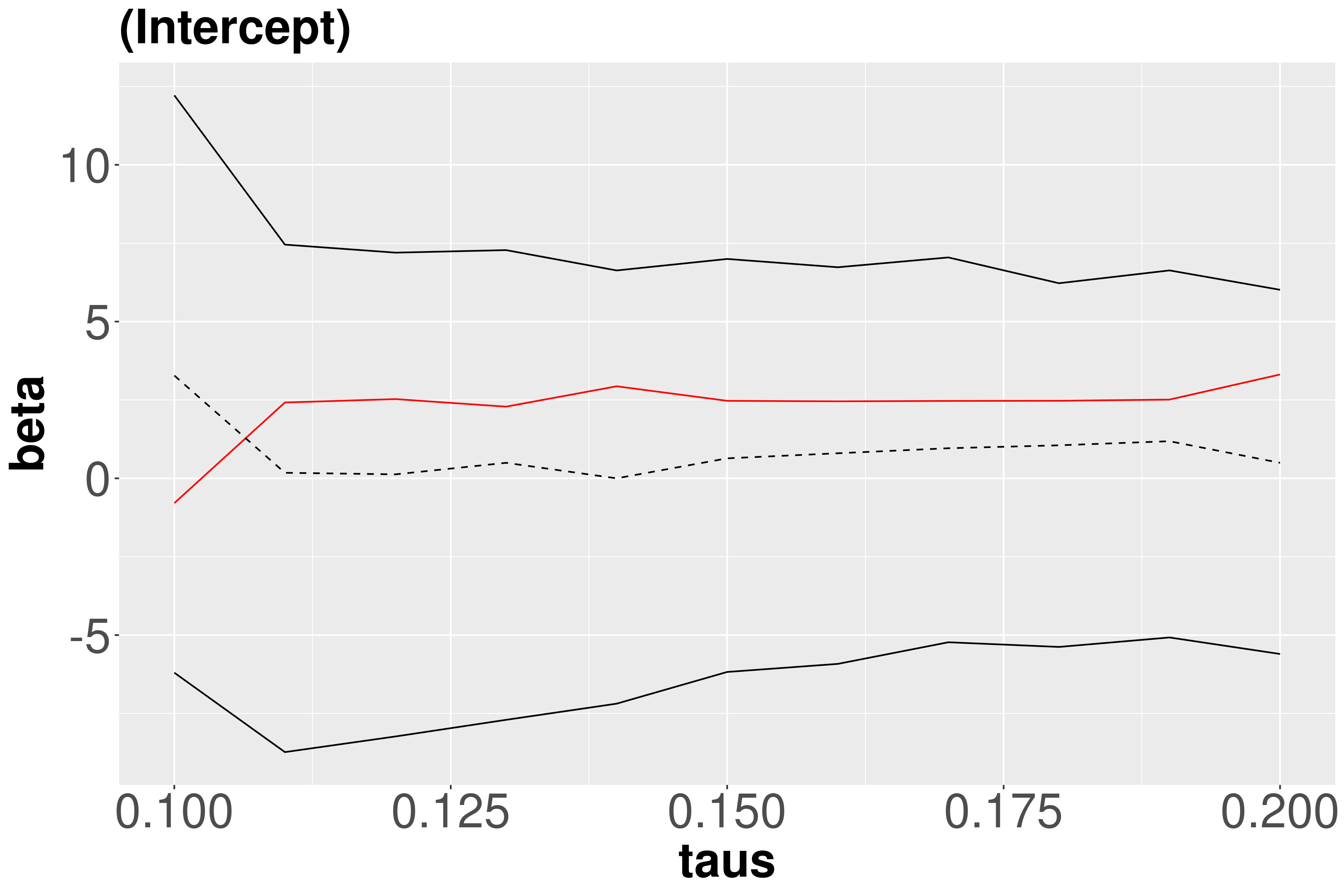}}
\subfigure[]{\label{fig:b}\includegraphics[width=80mm]{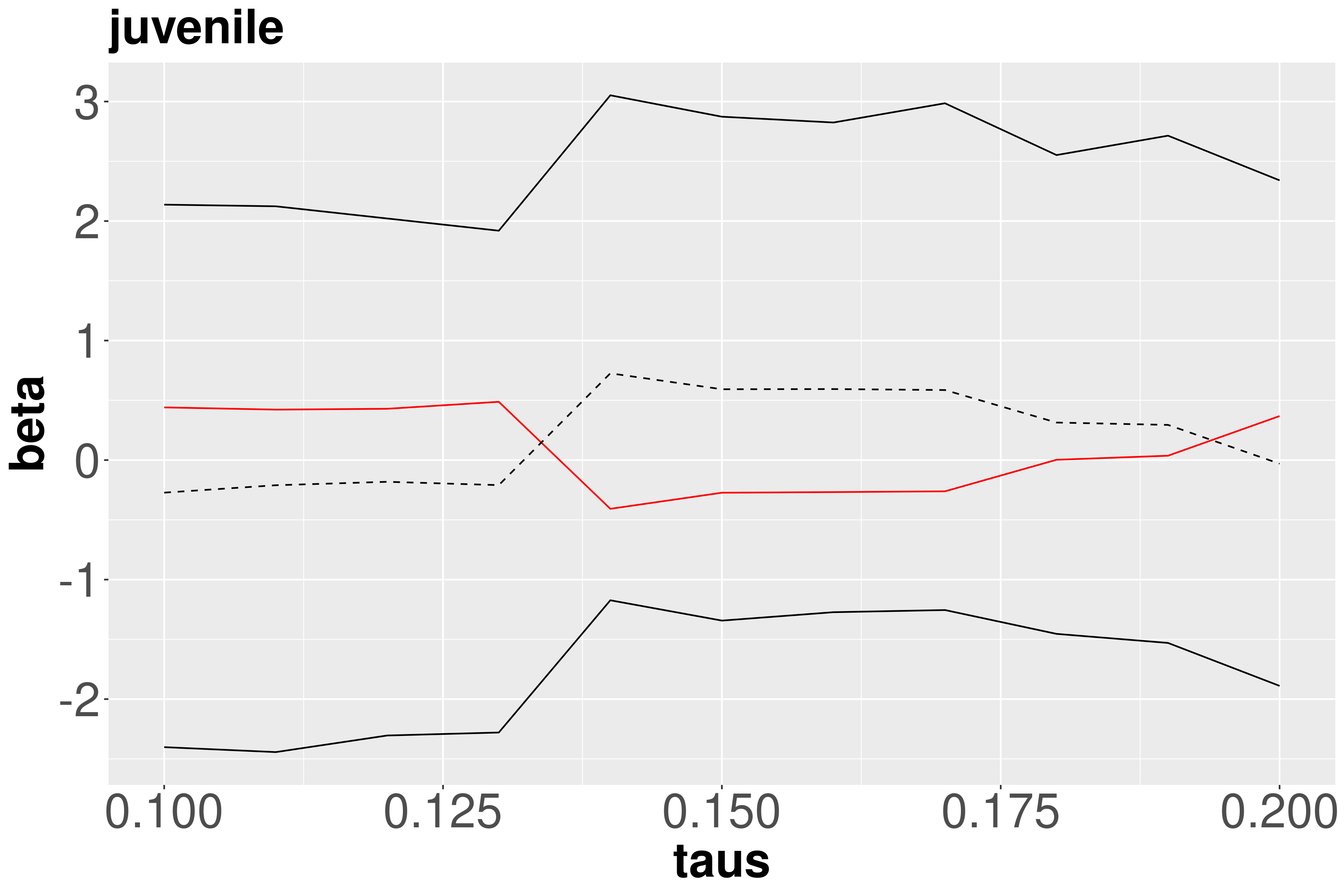}}
\subfigure[]{\includegraphics[width=80mm]{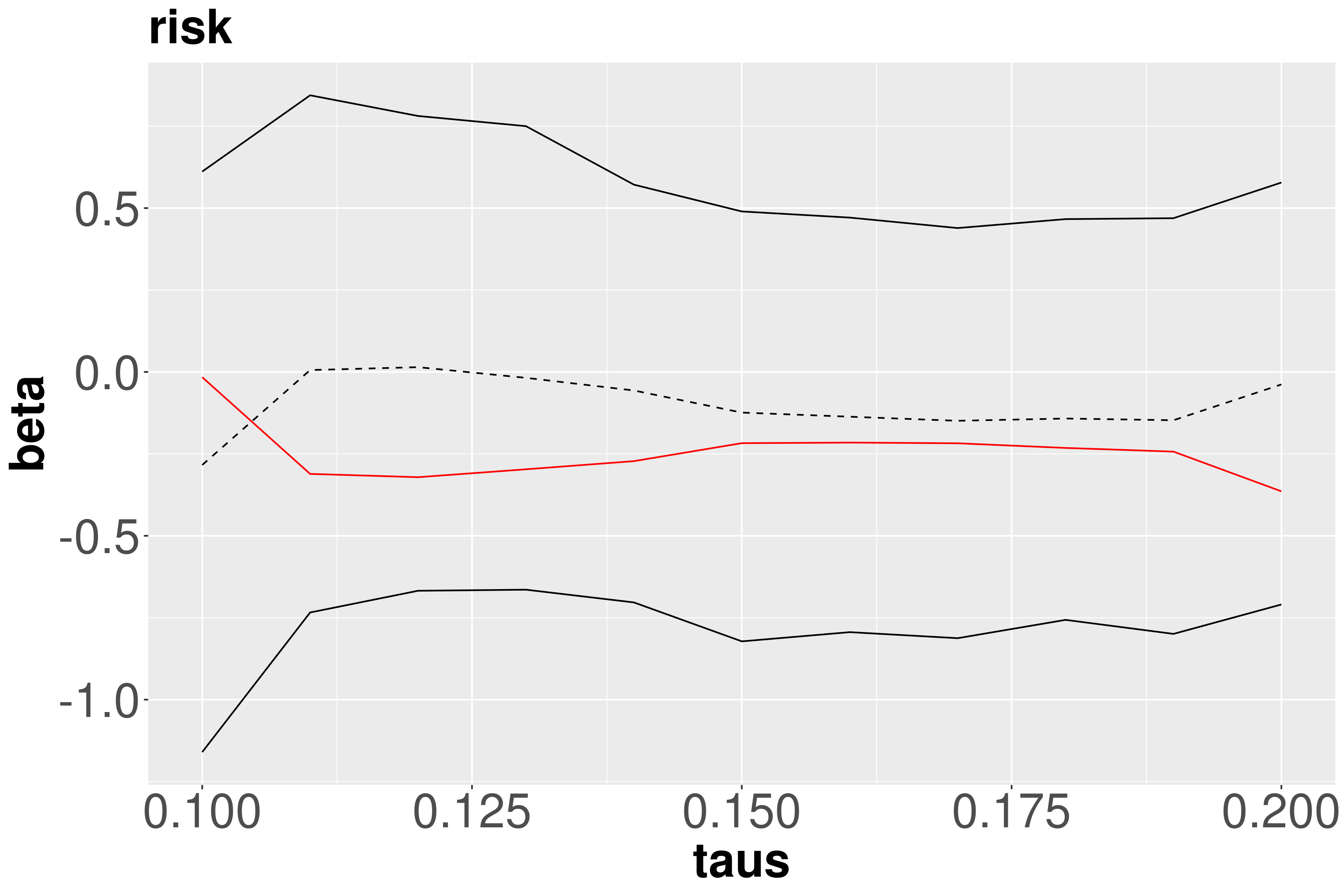}}
\caption{\label{fig:cov4}  The process $\{\hat\beta_{\text{argon}} - \hat\beta_{\text{xenon}}\}(\tau)$ for each coefficient (in red), together with $2.5\%$ and $97.5\%$ percentiles of the distribution of the bootstrap process 
$\{( \hat\beta_{\text{argon}}^* - \hat\beta_{\text{xenon}}^*) - (\hat\beta_{\text{argon}} - \hat\beta_{\text{xenon}}) \}(\tau)$ (black solid lines) and the mean bootstrap process (dotted black line). Subfigures (a), (b) and (c) show the estimated intercept and the estimated coefficients of \textit{juvenile} and \textit{risk} respectively, for a range of $\tau$-values.  }
\end{figure}

Finally, we can verify if there is a difference in quantiles between the treatment with the \textit{argon} and the treatment with the \textit{xenon} laser. For that purpose, we consider only the data regarding the treated eyes. There are $n_{\text{argon}} = 100$ patients who received the treatment with the \textit{argon} laser and $n_{\text{xenon}} = 97$ who were treated with the \textit{xenon} laser. For such comparison we use our method in the framework of independent samples (with different sample sizes) and test $H_0: \beta_{\text{argon}}(\tau) = \beta_{\text{xenon}}(\tau)$ for $\tau\in A$, including the same covariate variables as before. In order to avoid undetermined values, we set $A = [0.1, 0.2]$, and step size equal to 0.01. The number of bootstrap resamples and the choice of the estimator is the same as before. 
We include graphical and numerical results of the estimators in Figure \ref{fig:cov4} and Table \ref{table:covcoeff4}, respectively. For any choice of the test statistic the  p-value is greater than 0.05. Therefore, we conclude that no significant difference (at level of confidence 0.95) can be revealed between the quantiles for the two types of laser treatments. These results agree with specific studies on laser treatments reported, for instance, in \cite{plumb1982comparative}.

\begin{table}[H]
\centering
\begin{tabular}{|c|ccc|}
\hline
            & \multicolumn{1}{c|}{$\tau$= 0.10} & \multicolumn{1}{c|}{$\tau$ = 0.15} & $\tau$ = 0.20 \\ \hline
            & \multicolumn{3}{c|}{$\hat\beta_{\text{argon}}$}                                        \\ \hline
intercept & \multicolumn{1}{c|}{2.57}      & \multicolumn{1}{c|}{6.11}      & 5.92      \\ \hline
juvenile    & \multicolumn{1}{c|}{-0.48}     & \multicolumn{1}{c|}{-0.94}     & -0.82     \\ \hline
risk        & \multicolumn{1}{c|}{-0.03}     & \multicolumn{1}{c|}{-0.25}     & -0.23     \\ \hline
            & \multicolumn{3}{c|}{$\hat\beta_{\text{xenon}}$}                                          \\ \hline
intercept & \multicolumn{1}{c|}{3.36}      & \multicolumn{1}{c|}{3.64}      & 2.60      \\ \hline
juvenile    & \multicolumn{1}{c|}{-0.92}     & \multicolumn{1}{c|}{-0.66}     & -1.18     \\ \hline
risk        & \multicolumn{1}{c|}{-0.01}     & \multicolumn{1}{c|}{-0.04}     & 0.14      \\ \hline
            &  \multicolumn{3}{c|}{$\hat\beta_{\text{argon}}-\hat\beta_{\text{xenon}}$}                             \\ \hline
intercept & \multicolumn{1}{c|}{-0.79}     & \multicolumn{1}{c|}{2.47}      & 3.31      \\ \hline
juvenile    & \multicolumn{1}{c|}{0.44}      & \multicolumn{1}{c|}{-0.27}     & 0.37      \\ \hline
risk        & \multicolumn{1}{c|}{-0.02}     & \multicolumn{1}{c|}{-0.22}     & -0.36     \\ \hline
\end{tabular}
\caption{\label{table:covcoeff4}
Values of the components intercept, \textit{juvenile} and \textit{risk} of the processes   $\hat\beta_{\text{argon}}(\tau)$, $\hat\beta_{\text{xenon}}(\tau)$ and their difference, for $\tau\in\{0.10,0.15,0.20\}$.}
\end{table}

\section{Conclusions} \label{sec8}

This paper contributes to the literature regarding the comparison of two populations in the presence of covariates, by allowing the response to be right censored and by comparing the two populations by means of quantile curves, which are supposed to be linear in the covariates.  Both the case of independent samples and of paired samples is considered.  This setup has never been considered before in the literature, although it has many applications in practice.   

The paper can be extended in many directions. First of all, it would be interesting to extend the methodology to the case where there are more than two populations.   This extension should be possible, although definitely not straightforward.   See e.g.\ \cite{pardo2006comparison}, who considers the case of more than two populations in a different context.  Second, we did not use all existing quantile regression estimators in our simulation study.  A thorough simulation study in which all existing estimators are compared, would shed light on the relative performance of the tests based on these estimators.  And finally, the extension to other types of incomplete data would also be a useful development.   



\bibliographystyle{dcu} 
\bibliography{references}
\newpage
\appendix

\section{Proofs} 

\noindent
{\bf Proof of Theorem \ref{theo1}.}  From assumption (A) and the Donsker theorem we know that $n_j^{1/2}(\hat{\beta}_j - \beta_j)(\tau)$ converges weakly to a tight Gaussian process $G_j(\tau)$ with mean $0$ and covariance function given by $\Sigma_j(\tau_1,\tau_2) = E[g_j(\tau_1,X_j,\Delta_j,Z_j) g_j(\tau_2,X_j,\Delta_j,Z_j)^T]$.   
Then, by Slutsky's theorem and assumption (EG), we have that 
$n^{1/2}(\hat{\beta}_j - \beta_j)(\tau)$ converges weakly to $p_j^{-1/2} G_j(\tau)$, and so the limiting covariance function is given by $p_j^{-1}\Sigma_j(\tau_1,\tau_2)$. Finally, from the independence of the two samples, we can conclude that under $H_0$, $n^{1/2}(\hat{\beta}_1 - \hat{\beta}_2)(\tau) = n^{1/2}(\hat{\beta}_1 - \beta_1)(\tau) - n^{1/2}(\hat{\beta}_2 - \beta_2)(\tau)$ converges weakly to $p_1^{-1/2} G_1(\tau) - p_2^{-1/2} G_2(\tau)$, which is a tight Gaussian process with mean $0$ and covariance function given by $p_1^{-1}\Sigma_1(\tau_1,\tau_2) + p_2^{-1}\Sigma_2(\tau_1,\tau_2)$.  \hfill $\Box$ \\

\noindent
{\bf Proof of Corollary \ref{cor2}.} Under the local alternative $H_1$, which takes also the form $\beta_2(\tau) = \beta_1(\tau) - n^{-1/2}b(\tau)$, the process $ n^{1/2}(\hat{\beta}_1 - \hat{\beta}_2)(\tau)$ can be written as $ n^{1/2}(\hat{\beta}_1 - \beta_1)(\tau) - n^{1/2}(\hat{\beta}_2 - \beta_2)(\tau) + b(\tau)$. Since $b(\tau)$ is deterministic, the limiting process is a non-centred Gaussian process with covariance function given as in Theorem \ref{theo1}. This means that 
$$ n^{1/2}(\hat{\beta}_1 - \hat{\beta}_2)(\tau)\xrightarrow{}_d W_I(\tau) + b(\tau),$$
where $W_I$ is the same Gaussian process as in Theorem \ref{theo1}. Since $T_{L_2}$, $T_{L_\infty}$ and $T_{B,k}$ are continuous maps, by the continuous mapping theorem we obtain the result. \hfill $\Box$ \\

\noindent
{\bf Proof of Theorem \ref{theo2}.}  From assumption (A) it follows that under $H_0$,
\begin{align*}
& n^{1/2} (\hat{\beta}_1(\tau) - \hat{\beta}_2(\tau)) \\
& = n^{1/2} (\hat{\beta}_1(\tau) - \beta_1(\tau)) -  n^{1/2} (\hat{\beta}_2(\tau) - \beta_2(\tau)) \\
& = n^{-1/2} \sum_{i=1}^n [g_1(\tau,X_{1i},\Delta_{1i},Z_{1i}) - g_2(\tau,X_{2i},\Delta_{2i},Z_{2i})] + o_P(1),
\end{align*}
uniformly in $\tau \in A$.   Since the difference of two Donsker classes is Donsker (see Example 2.10.7 in \cite{Vaart1996}), we can conclude that $n^{1/2} (\hat{\beta}_1(\tau) - \hat{\beta}_2(\tau))$ converges weakly to a zero-mean Gaussian process with covariance function given in the statement of the theorem.   \hfill $\Box$ \\

\noindent
{\bf Proof of Theorem \ref{theo3}.} 
First we consider the two samples separately.
By classical results on empirical process theory (see e.g.\ Theorem 3.2 in \cite{gine1996empirical}), condition (A$^*$) assures that $n^{1/2} (\hat\beta_  {j}^* - \hat\beta_{j})$ converges to a tight Gaussian process and moreover, since $\eta_{ji}$ has mean zero and variance equal to 1, the asymptotic covariance operator of the process $n^{1/2} (\hat\beta_j^* - \hat\beta_j)$  conditional on the data equals
\begin{align*}
& E(\eta_j^2) \frac{n}{n_j^2} \sum_{i=1}^{n_j} g_j(\tau_1,X_{ji},\Delta_{ji},Z_{ji}) g_j(\tau_2,X_{ji},\Delta_{ji},Z_{ji})^T \\
& \stackrel{a.s.}{\rightarrow} p_j^{-1} E[g_j(\tau_1,X_j,\Delta_j,Z_j) g_j(\tau_2,X_j,\Delta_j,Z_j)^T],
\end{align*}
and this is the asymptotic covariance of the process $n^{1/2} (\hat\beta_j - \beta_j)$.

The extension to two samples is similar to what was done in the proof of Theorem \ref{theo1} for independent samples and Theorem \ref{theo2} for paired samples.  The two processes to consider are now $n^{1/2}(\hat\beta_j^* - \hat\beta_j)(\tau)$ for $j = 1,2$.  Note that in the case of independent samples, the two bootstrap processes are independent, since the $\eta_{ji}$ are all i.i.d.  It is therefore straightforward to show that the asymptotic covariance of $n^{1/2}(\hat\beta_1^*(\tau)-\hat\beta_2^*(\tau) - \hat\beta_1(\tau)+\hat\beta_2(\tau) )$ converges to the one of the original process.  In the case of paired samples instead, the bootstrap asymptotic covariance conditional on the data equals (with $V_{ji} = (X_{ji},\Delta_{ji},Z_{ji})$, $j=1,2$)
\begin{align*}
& E(\eta_1^2) \frac{1}{n} \sum_{i=1}^n [g_1(\tau_1,V_{1i}) - g_2(\tau_1,V_{2i})] 
[g_1(\tau_2,V_{1i}) - g_2(\tau_2,V_{2i})]^T \\
& \stackrel{a.s.}{\rightarrow} E\{[g_1(\tau_1,V_1)-g_2(\tau_1,V_2)] [g_1(\tau_2,V_1)-g_2(\tau_2,V_2)]^T\},
\end{align*}
and this is the asymptotic covariance of $n^{1/2} (\hat\beta_1 - \hat\beta_2 - \beta_1 + \beta_2)$, from which the assertion follows. \hfill $\Box$ \\

\section{Simulation Results}

We include additional simulation results for 40\% of censoring.

\begin{table}[H]
\centering
\begin{adjustbox}{width=1\textwidth}
\small
\begin{tabular}{|c|c|c|ccc|ccc|ccc|ccc|c|} 
\cline{4-16}
\multicolumn{3}{c|}{\multirow{2}{*}{}} & Bonf. & $T_{L_2}$ & $T_{L_\infty}$ & Bonf. & $T_{L_2}$ & $T_{L_\infty}$ & Bonf. & $T_{L_2}$ & $T_{L_\infty}$ & Bonf. & $T_{L_2}$ & $T_{L_\infty}$ &        \\ 
\hline
Mod.                 & $n_2$                & Diff. &   \multicolumn{3}{c|}{\cite{debacker2019}}   & \multicolumn{3}{c|}{\cite{peng2008survival}} & \multicolumn{3}{c|}{\cite{portnoy2003censored}}   & \multicolumn{3}{c|}{\cite{wang2009locally}}       & SA     \\ 
\hline
\multirow{9}{*}{1}   & \multirow{3}{*}{100} & 0     & 0.038 & 0.014     & 0.020          & 0.034 & 0.022     & 0.024          & 0.010 & 0.006     & 0.004          & 0.030 & 0.018     & 0.022                               & 0.066  \\
                     &                      & 0.2   & 0.118 & 0.110     & 0.122          & 0.135 & 0.131     & 0.110          & 0.039 & 0.031     & 0.033          & 0.110 & 0.064     & 0.092                               & 0.148  \\
                     &                      & 0.4   & 0.506 & 0.502     & 0.456          & 0.586 & 0.609     & 0.582          & 0.433 & 0.435     & 0.352          & 0.570 & 0.568     & 0.524                               & 0.516  \\ 
\cline{2-16}
                     & \multirow{3}{*}{200} & 0     & 0.046 & 0.038     & 0.036          & 0.050 & 0.042     & 0.034          & 0.028 & 0.024     & 0.024          & 0.048 & 0.026     & 0.028                               & 0.032  \\
                     &                      & 0.2   & 0.248 & 0.204     & 0.238          & 0.236 & 0.212     & 0.200          & 0.180 & 0.182     & 0.160          & 0.220 & 0.224     & 0.218                               & 0.254  \\
                     &                      & 0.4   & 0.772 & 0.824     & 0.740          & 0.836 & 0.857     & 0.859          & 0.764 & 0.810     & 0.762          & 0.828 & 0.876     & 0.786                               & 0.702  \\ 
\cline{2-16}
                     & \multirow{3}{*}{400} & 0     & 0.046 & 0.044     & 0.038          & 0.030 & 0.022     & 0.020          & 0.018 & 0.002     & 0.006          & 0.032 & 0.018     & 0.026                               & 0.022  \\
                     &                      & 0.2   & 0.376 & 0.366     & 0.348          & 0.298 & 0.292     & 0.286          & 0.124 & 0.086     & 0.080          & 0.292 & 0.270     & 0.252                               & 0.358  \\
                     &                      & 0.4   & 0.932 & 0.952     & 0.904          & 0.878 & 0.910     & 0.880          & 0.762 & 0.830     & 0.782          & 0.882 & 0.906     & 0.884                               & 0.792  \\ 
\hline
\multirow{9}{*}{2}   & \multirow{3}{*}{100} & 0     & 0.032 & 0.026     & 0.024          & 0.030 & 0.020     & 0.024          & 0.024 & 0.008     & 0.012          & 0.036 & 0.004     & 0.018                               & 0.032  \\
                     &                      & 0.2   & 0.216 & 0.230     & 0.184          & 0.362 & 0.284     & 0.270          & 0.112 & 0.137     & 0.102          & 0.204 & 0.172     & 0.160                               & 0.106  \\
                     &                      & 0.4   & 0.888 & 0.938     & 0.896          & 0.888 & 0.914     & 0.876          & 0.830 & 0.865     & 0.804          & 0.884 & 0.892     & 0.856                               & 0.556  \\ 
\cline{2-16}
                     & \multirow{3}{*}{200} & 0     & 0.060 & 0.034     & 0.038          & 0.038 & 0.028     & 0.022          & 0.038 & 0.036     & 0.026          & 0.034 & 0.028     & 0.032                               & 0.050  \\
                     &                      & 0.2   & 0.556 & 0.446     & 0.454          & 0.470 & 0.450     & 0.396          & 0.442 & 0.476     & 0.422          & 0.458 & 0.426     & 0.402                               & 0.222  \\
                     &                      & 0.4   & 0.998 & 0.996     & 0.994          & 0.984 & 0.998     & 0.990          & 0.986 & 0.990     & 0.986          & 0.978 & 0.990     & 0.982                               & 0.752  \\ 
\cline{2-16}
                     & \multirow{3}{*}{400} & 0     & 0.036 & 0.022     & 0.026          & 0.056 & 0.046     & 0.048          & 0.018 & 0.008     & 0.010          & 0.060 & 0.036     & 0.038                               & 0.050  \\
                     &                      & 0.2   & 0.564 & 0.684     & 0.638          & 0.640 & 0.642     & 0.618          & 0.474 & 0.464     & 0.436          & 0.596 & 0.662     & 0.578                               & 0.350  \\
                     &                      & 0.4   & 0.998 & 1.000     & 1.000          & 1.000 & 1.000     & 1.000          & 0.996 & 0.996     & 0.994          & 1.000 & 1.000     & 1.000                               & 0.844  \\ 
\hline
\multirow{9}{*}{3}   & \multirow{3}{*}{100} & 0     & 0.016 & 0.012     & 0.016          & 0.036 & 0.016     & 0.036          & 0.016 & 0.004     & 0.012          & 0.030 & 0.018     & 0.026                               & 0.048  \\
                     &                      & 0.2   & 0.476 & 0.470     & 0.480          & 0.512 & 0.510     & 0.510          & 0.261 & 0.261     & 0.180          & 0.520 & 0.538     & 0.530                               & 0.162  \\
                     &                      & 0.4   & 0.996 & 0.998     & 0.996          & 0.996 & 1.000     & 1.000          & 0.982 & 0.990     & 0.978          & 1.000 & 1.000     & 1.000                               & 0.752  \\ 
\cline{2-16}
                     & \multirow{3}{*}{200} & 0     & 0.032 & 0.028     & 0.026          & 0.046 & 0.026     & 0.024          & 0.028 & 0.024     & 0.026          & 0.034 & 0.030     & 0.026                               & 0.046  \\
                     &                      & 0.2   & 0.790 & 0.796     & 0.762          & 0.774 & 0.780     & 0.736          & 0.664 & 0.754     & 0.674          & 0.748 & 0.772     & 0.740                               & 0.306  \\
                     &                      & 0.4   & 1.000 & 1.000     & 1.000          & 1.000 & 1.000     & 1.000          & 1.000 & 1.000     & 1.000          & 0.998 & 1.000     & 1.000                               & 0.934  \\ 
\cline{2-16}
                     & \multirow{3}{*}{400} & 0     & 0.052 & 0.034     & 0.038          & 0.048 & 0.030     & 0.030          & 0.016 & 0.008     & 0.006          & 0.036 & 0.018     & 0.016                               & 0.040  \\
                     &                      & 0.2   & 0.910 & 0.916     & 0.892          & 0.918 & 0.926     & 0.916          & 0.716 & 0.794     & 0.752          & 0.878 & 0.874     & 0.860                               & 0.490  \\
                     &                      & 0.4   & 1.000 & 1.000     & 1.000          & 1.000 & 1.000     & 1.000          & 1.000 & 1.000     & 1.000          & 1.000 & 1.000     & 1.000                               & 0.970  \\
\hline
\end{tabular}
\end{adjustbox}
\caption{\label{table:independentDistribution40}Simulation results for the comparison of quantile curves (range of $\tau$-values) for two independent samples for percentage of censoring equal to 40\%.  The size of the first sample is $n_1=200$, the significance level is $\alpha=0.05$ and Diff $=\beta_{21}-\beta_{11}$, so the null hypothesis is satisfied for Diff $=0$. The abbreviation SA stands for the test statistic of \cite{sant2014nonparametric}, and Bonf, $T_{L_2}$ and $T_{L_\infty}$ stand for the test statistics defined in (\ref{Bonf}), (\ref{L2}) and (\ref{Linfty}), respectively. }
\end{table}

\begin{table}[H]
\centering
\begin{adjustbox}{width=1\textwidth}
\small
\begin{tabular}{|c|c|c|ccc|ccc|ccc|ccc|} 
\cline{4-15}
\multicolumn{1}{c}{} & \multicolumn{1}{c}{} &       & Bonf  & $T_{L_2}$ & $T_{L_\infty}$ & Bonf  & $T_{L_2}$ & $T_{L_\infty}$ & Bonf  & $T_{L_2}$ & $T_{L_\infty}$ & Bonf  & $T_{L_2}$ & $T_{L_\infty}$  \\ 
\hline
Mod.                 & $n_2$                & Diff. & \multicolumn{3}{c|}{\cite{debacker2019}} & \multicolumn{3}{c|}{\cite{peng2008survival}} & \multicolumn{3}{c|}{\cite{portnoy2003censored}}   & \multicolumn{3}{c|}{\cite{wang2009locally}}   \\ 
\hline
\multirow{9}{*}{1}   & \multirow{3}{*}{100} & 0     & 0.016 & 0.008   & 0.014        & 0.028 & 0.022   & 0.030        & 0.004 & 0.004   & 0.004        & 0.020 & 0.018   & 0.022         \\
                     &                      & 0.2   & 0.054 & 0.070   & 0.056        & 0.044 & 0.044   & 0.034        & 0.014 & 0.016   & 0.016        & 0.040 & 0.030   & 0.040         \\
                     &                      & 0.4   & 0.344 & 0.358   & 0.316        & 0.200 & 0.234   & 0.206        & 0.202 & 0.166   & 0.200        & 0.234 & 0.288   & 0.262         \\ 
\cline{2-15}
                     & \multirow{3}{*}{200} & 0     & 0.022 & 0.024   & 0.020        & 0.018 & 0.012   & 0.018        & 0.020 & 0.012   & 0.020        & 0.024 & 0.016   & 0.024         \\
                     &                      & 0.2   & 0.118 & 0.146   & 0.106        & 0.108 & 0.122   & 0.106        & 0.122 & 0.104   & 0.118        & 0.080 & 0.082   & 0.076         \\
                     &                      & 0.4   & 0.640 & 0.694   & 0.628        & 0.470 & 0.516   & 0.458        & 0.586 & 0.616   & 0.600        & 0.530 & 0.586   & 0.442         \\ 
\cline{2-15}
                     & \multirow{3}{*}{400} & 0     & 0.010 & 0.014   & 0.012        & 0.012 & 0.020   & 0.014        & 0.010 & 0.014   & 0.010        & 0.004 & 0.004   & 0.006         \\
                     &                      & 0.2   & 0.252 & 0.288   & 0.264        & 0.122 & 0.122   & 0.110        & 0.068 & 0.058   & 0.068        & 0.168 & 0.182   & 0.150         \\
                     &                      & 0.4   & 0.660 & 0.828   & 0.718        & 0.674 & 0.772   & 0.676        & 0.498 & 0.666   & 0.488        & 0.756 & 0.826   & 0.756         \\ 
\hline
\multirow{9}{*}{2}   & \multirow{3}{*}{100} & 0     & 0.024 & 0.018   & 0.024        & 0.010 & 0.010   & 0.010        & 0.004 & 0.002   & 0.006        & 0.028 & 0.022   & 0.028         \\
                     &                      & 0.2   & 0.086 & 0.102   & 0.088        & 0.084 & 0.114   & 0.086        & 0.068 & 0.078   & 0.074        & 0.114 & 0.144   & 0.116         \\
                     &                      & 0.4   & 0.754 & 0.842   & 0.800        & 0.566 & 0.658   & 0.606        & 0.569 & 0.645   & 0.577        & 0.736 & 0.804   & 0.744         \\ 
\cline{2-15}
                     & \multirow{3}{*}{200} & 0     & 0.028 & 0.014   & 0.024        & 0.030 & 0.028   & 0.026        & 0.026 & 0.022   & 0.024        & 0.022 & 0.014   & 0.016         \\
                     &                      & 0.2   & 0.326 & 0.380   & 0.286        & 0.232 & 0.266   & 0.222        & 0.326 & 0.360   & 0.308        & 0.306 & 0.328   & 0.284         \\
                     &                      & 0.4   & 0.952 & 0.980   & 0.954        & 0.892 & 0.944   & 0.880        & 0.930 & 0.962   & 0.926        & 0.812 & 0.914   & 0.828         \\ 
\cline{2-15}
                     & \multirow{3}{*}{400} & 0     & 0.022 & 0.024   & 0.022        & 0.022 & 0.010   & 0.022        & 0.002 & 0.002   & 0.000        & 0.026 & 0.036   & 0.026         \\
                     &                      & 0.2   & 0.418 & 0.468   & 0.424        & 0.310 & 0.450   & 0.368        & 0.224 & 0.296   & 0.232        & 0.294 & 0.450   & 0.366         \\
                     &                      & 0.4   & 0.992 & 0.998   & 0.988        & 0.984 & 0.996   & 0.984        & 0.948 & 0.976   & 0.940        & 0.974 & 0.996   & 0.982         \\ 
\hline
\multirow{9}{*}{3}   & \multirow{3}{*}{100} & 0     & 0.010 & 0.004   & 0.010        & 0.024 & 0.016   & 0.020        & 0.000 & 0.000   & 0.000        & 0.012 & 0.014   & 0.012         \\
                     &                      & 0.2   & 0.342 & 0.394   & 0.322        & 0.273 & 0.333   & 0.248        & 0.126 & 0.106   & 0.118        & 0.214 & 0.276   & 0.210         \\
                     &                      & 0.4   & 0.960 & 0.970   & 0.962        & 0.958 & 0.980   & 0.954        & 0.870 & 0.940   & 0.868        & 0.910 & 0.966   & 0.930         \\ 
\cline{2-15}
                     & \multirow{3}{*}{200} & 0     & 0.030 & 0.028   & 0.024        & 0.026 & 0.020   & 0.022        & 0.022 & 0.012   & 0.020        & 0.042 & 0.028   & 0.050         \\
                     &                      & 0.2   & 0.514 & 0.628   & 0.506        & 0.504 & 0.572   & 0.512        & 0.440 & 0.612   & 0.482        & 0.472 & 0.622   & 0.472         \\
                     &                      & 0.4   & 0.996 & 0.998   & 0.994        & 0.998 & 1.000   & 0.998        & 0.998 & 1.000   & 0.998        & 0.998 & 1.000   & 0.996         \\ 
\cline{2-15}
                     & \multirow{3}{*}{400} & 0     & 0.014 & 0.012   & 0.016        & 0.006 & 0.004   & 0.006        & 0.004 & 0.008   & 0.004        & 0.008 & 0.010   & 0.008         \\
                     &                      & 0.2   & 0.660 & 0.724   & 0.648        & 0.716 & 0.750   & 0.704        & 0.512 & 0.526   & 0.474        & 0.672 & 0.754   & 0.632         \\
                     &                      & 0.4   & 1.000 & 1.000   & 1.000        & 1.000 & 1.000   & 1.000        & 1.000 & 1.000   & 1.000        & 1.000 & 1.000   & 1.000         \\
\hline
\end{tabular}
\end{adjustbox}
\caption{\label{table:independentMedian40}Simulation results for the comparison of the median ($\tau=0.5$) for two independent samples subject to $40\%$ censoring. The size of the first sample is $n_1=200$, the significance level is $\alpha=0.05$ and Diff $=\beta_{21}-\beta_{11}$, so the null hypothesis is satisfied for Diff $=0$. The abbreviations Bonf, $T_{L_2}$ and $T_{L_\infty}$ stand for the test statistics defined in (\ref{Bonf}), (\ref{L2}) and (\ref{Linfty}), respectively.  }
\end{table}

\begin{table}[H]
\centering
\begin{adjustbox}{totalheight=\textheight-3\baselineskip}
\small
\begin{tabular}{|c|c|c|ccc|ccc|ccc|ccc|} 
\cline{4-15}
\multicolumn{1}{c}{} & \multicolumn{1}{c}{} &       & Bonf  & $T_{L_2}$ & $T_{L_\infty}$ & Bonf  & $T_{L_2}$ & $T_{L_\infty}$ & Bonf  & $T_{L_2}$ & $T_{L_\infty}$ & Bonf  & $T_{L_2}$ & $T_{L_\infty}$  \\ 
\cline{4-15}
\multicolumn{1}{c}{} & \multicolumn{1}{c}{} &       & \multicolumn{3}{c|}{\cite{debacker2019}}     & \multicolumn{3}{c|}{\cite{peng2008survival}}     & \multicolumn{3}{c|}{\cite{portnoy2003censored}}       & \multicolumn{3}{c|}{\cite{wang2009locally}}       \\ 
\hline
Mod.                 & $n$                    & Diff. & \multicolumn{12}{c|}{$\eta = 0.2$}                                                                                                                 \\ 
\hline
\multirow{9}{*}{1}   & \multirow{3}{*}{100} & 0     & 0.026 & 0.018     & 0.016          & 0.013 & 0.007     & 0.009          & 0.026 & 0.011     & 0.009          & 0.044 & 0.014     & 0.018           \\
                     &                      & 0.2   & 0.098 & 0.058     & 0.080          & 0.108 & 0.068     & 0.066          & 0.069 & 0.041     & 0.037          & 0.090 & 0.046     & 0.058           \\
                     &                      & 0.4   & 0.196 & 0.204     & 0.184          & 0.156 & 0.098     & 0.111          & 0.245 & 0.182     & 0.200          & 0.238 & 0.170     & 0.172           \\ 
\cline{2-15}
                     & \multirow{3}{*}{200} & 0     & 0.032 & 0.018     & 0.018          & 0.051 & 0.018     & 0.028          & 0.047 & 0.029     & 0.029          & 0.034 & 0.014     & 0.016           \\
                     &                      & 0.2   & 0.118 & 0.128     & 0.108          & 0.119 & 0.092     & 0.098          & 0.095 & 0.082     & 0.086          & 0.088 & 0.072     & 0.078           \\
                     &                      & 0.4   & 0.428 & 0.468     & 0.386          & 0.483 & 0.496     & 0.448          & 0.383 & 0.358     & 0.286          & 0.370 & 0.382     & 0.358           \\ 
\cline{2-15}
                     & \multirow{3}{*}{400} & 0     & 0.038 & 0.032     & 0.032          & 0.042 & 0.018     & 0.014          & 0.032 & 0.014     & 0.008          & 0.036 & 0.026     & 0.022           \\
                     &                      & 0.2   & 0.202 & 0.212     & 0.198          & 0.291 & 0.222     & 0.194          & 0.232 & 0.222     & 0.222          & 0.276 & 0.224     & 0.226           \\
                     &                      & 0.4   & 0.820 & 0.842     & 0.816          & 0.802 & 0.838     & 0.796          & 0.831 & 0.880     & 0.835          & 0.750 & 0.756     & 0.732           \\ 
\hline
\multirow{9}{*}{2}   & \multirow{3}{*}{100} & 0     & 0.024 & 0.010     & 0.016          & 0.030 & 0.015     & 0.017          & 0.025 & 0.021     & 0.025          & 0.018 & 0.008     & 0.006           \\
                     &                      & 0.2   & 0.080 & 0.036     & 0.044          & 0.077 & 0.060     & 0.045          & 0.048 & 0.035     & 0.050          & 0.084 & 0.040     & 0.048           \\
                     &                      & 0.4   & 0.274 & 0.292     & 0.236          & 0.375 & 0.330     & 0.292          & 0.336 & 0.351     & 0.347          & 0.261 & 0.246     & 0.204           \\ 
\cline{2-15}
                     & \multirow{3}{*}{200} & 0     & 0.036 & 0.030     & 0.026          & 0.050 & 0.018     & 0.018          & 0.026 & 0.024     & 0.030          & 0.030 & 0.026     & 0.024           \\
                     &                      & 0.2   & 0.230 & 0.258     & 0.202          & 0.240 & 0.236     & 0.192          & 0.160 & 0.148     & 0.130          & 0.200 & 0.168     & 0.130           \\
                     &                      & 0.4   & 0.808 & 0.778     & 0.770          & 0.670 & 0.796     & 0.686          & 0.782 & 0.739     & 0.731          & 0.748 & 0.732     & 0.702           \\ 
\cline{2-15}
                     & \multirow{3}{*}{400} & 0     & 0.042 & 0.024     & 0.020          & 0.052 & 0.048     & 0.042          & 0.036 & 0.024     & 0.024          & 0.034 & 0.020     & 0.024           \\
                     &                      & 0.2   & 0.356 & 0.362     & 0.354          & 0.533 & 0.561     & 0.509          & 0.511 & 0.511     & 0.481          & 0.554 & 0.486     & 0.462           \\
                     &                      & 0.4   & 0.992 & 0.992     & 0.992          & 0.982 & 0.990     & 0.984          & 0.982 & 0.986     & 0.984          & 0.996 & 0.996     & 0.990           \\ 
\hline
\multirow{9}{*}{3}   & \multirow{3}{*}{100} & 0     & 0.030 & 0.022     & 0.018          & 0.042 & 0.034     & 0.034          & 0.016 & 0.016     & 0.014          & 0.024 & 0.010     & 0.008           \\
                     &                      & 0.2   & 0.170 & 0.136     & 0.136          & 0.182 & 0.117     & 0.115          & 0.139 & 0.139     & 0.129          & 0.226 & 0.150     & 0.172           \\
                     &                      & 0.4   & 0.540 & 0.636     & 0.492          & 0.705 & 0.600     & 0.575          & 0.648 & 0.650     & 0.616          & 0.568 & 0.580     & 0.514           \\ 
\cline{2-15}
                     & \multirow{3}{*}{200} & 0     & 0.042 & 0.024     & 0.026          & 0.056 & 0.038     & 0.040          & 0.042 & 0.034     & 0.034          & 0.030 & 0.018     & 0.014           \\
                     &                      & 0.2   & 0.406 & 0.408     & 0.380          & 0.368 & 0.364     & 0.308          & 0.321 & 0.379     & 0.303          & 0.368 & 0.390     & 0.358           \\
                     &                      & 0.4   & 0.956 & 0.982     & 0.960          & 0.976 & 0.984     & 0.972          & 0.952 & 0.974     & 0.948          & 0.964 & 0.986     & 0.978           \\ 
\cline{2-15}
                     & \multirow{3}{*}{400} & 0     & 0.072 & 0.064     & 0.066          & 0.070 & 0.042     & 0.042          & 0.026 & 0.008     & 0.018          & 0.030 & 0.012     & 0.018           \\
                     &                      & 0.2   & 0.786 & 0.806     & 0.758          & 0.810 & 0.824     & 0.818          & 0.758 & 0.846     & 0.738          & 0.746 & 0.780     & 0.762           \\
                     &                      & 0.4   & 1.000 & 1.000     & 1.000          & 1.000 & 1.000     & 1.000          & 1.000 & 1.000     & 1.000          & 1.000 & 1.000     & 1.000           \\ 
\hline
\multicolumn{3}{|c|}{}        & \multicolumn{12}{c|}{$\eta = 0.4$}                                                                                                                 \\ 
\hline
\multirow{9}{*}{1}   & \multirow{3}{*}{100} & 0     & 0.046 & 0.022     & 0.016          & 0.023 & 0.011     & 0.006          & 0.017 & 0.004     & 0.011          & 0.012 & 0.004     & 0.008           \\
                     &                      & 0.2   & 0.090 & 0.058     & 0.060          & 0.079 & 0.051     & 0.056          & 0.035 & 0.012     & 0.021          & 0.086 & 0.058     & 0.064           \\
                     &                      & 0.4   & 0.234 & 0.230     & 0.184          & 0.257 & 0.161     & 0.176          & 0.215 & 0.185     & 0.180          & 0.226 & 0.158     & 0.146           \\ 
\cline{2-15}
                     & \multirow{3}{*}{200} & 0     & 0.036 & 0.032     & 0.032          & 0.044 & 0.036     & 0.038          & 0.052 & 0.047     & 0.043          & 0.020 & 0.010     & 0.016           \\
                     &                      & 0.2   & 0.162 & 0.144     & 0.148          & 0.116 & 0.089     & 0.081          & 0.164 & 0.130     & 0.134          & 0.124 & 0.102     & 0.110           \\
                     &                      & 0.4   & 0.656 & 0.604     & 0.582          & 0.518 & 0.471     & 0.438          & 0.693 & 0.663     & 0.633          & 0.502 & 0.494     & 0.490           \\ 
\cline{2-15}
                     & \multirow{3}{*}{400} & 0     & 0.030 & 0.022     & 0.020          & 0.052 & 0.030     & 0.042          & 0.036 & 0.018     & 0.022          & 0.028 & 0.016     & 0.022           \\
                     &                      & 0.2   & 0.392 & 0.306     & 0.320          & 0.359 & 0.399     & 0.347          & 0.395 & 0.393     & 0.391          & 0.330 & 0.274     & 0.256           \\
                     &                      & 0.4   & 0.918 & 0.936     & 0.910          & 0.921 & 0.943     & 0.921          & 0.931 & 0.951     & 0.945          & 0.926 & 0.950     & 0.924           \\ 
\cline{2-15}
\multirow{9}{*}{2}   & \multirow{3}{*}{100} & 0     & 0.008 & 0.008     & 0.008          & 0.019 & 0.008     & 0.012          & 0.032 & 0.022     & 0.028          & 0.024 & 0.012     & 0.012           \\
                     &                      & 0.2   & 0.090 & 0.058     & 0.044          & 0.117 & 0.077     & 0.092          & 0.092 & 0.058     & 0.066          & 0.092 & 0.056     & 0.042           \\
                     &                      & 0.4   & 0.530 & 0.458     & 0.386          & 0.471 & 0.348     & 0.342          & 0.464 & 0.457     & 0.365          & 0.449 & 0.377     & 0.315           \\ 
\cline{2-15}
                     & \multirow{3}{*}{200} & 0     & 0.048 & 0.020     & 0.022          & 0.032 & 0.036     & 0.016          & 0.024 & 0.020     & 0.020          & 0.066 & 0.022     & 0.030           \\
                     &                      & 0.2   & 0.312 & 0.228     & 0.230          & 0.236 & 0.234     & 0.201          & 0.242 & 0.202     & 0.169          & 0.276 & 0.206     & 0.176           \\
                     &                      & 0.4   & 0.890 & 0.950     & 0.896          & 0.918 & 0.924     & 0.892          & 0.912 & 0.929     & 0.904          & 0.940 & 0.912     & 0.880           \\ 
\cline{2-15}
                     & \multirow{3}{*}{400} & 0     & 0.036 & 0.016     & 0.016          & 0.048 & 0.008     & 0.022          & 0.048 & 0.032     & 0.032          & 0.042 & 0.030     & 0.034           \\
                     &                      & 0.2   & 0.582 & 0.562     & 0.524          & 0.588 & 0.606     & 0.560          & 0.673 & 0.707     & 0.669          & 0.584 & 0.598     & 0.548           \\
                     &                      & 0.4   & 1.000 & 1.000     & 1.000          & 1.000 & 1.000     & 1.000          & 1.000 & 1.000     & 1.000          & 1.000 & 1.000     & 0.998           \\ 
\cline{2-15}
\multirow{9}{*}{3}   & \multirow{3}{*}{100} & 0     & 0.020 & 0.014     & 0.012          & 0.028 & 0.006     & 0.006          & 0.022 & 0.006     & 0.010          & 0.022 & 0.012     & 0.012           \\
                     &                      & 0.2   & 0.214 & 0.178     & 0.172          & 0.181 & 0.161     & 0.159          & 0.226 & 0.188     & 0.162          & 0.168 & 0.132     & 0.132           \\
                     &                      & 0.4   & 0.820 & 0.814     & 0.746          & 0.819 & 0.832     & 0.799          & 0.754 & 0.813     & 0.770          & 0.710 & 0.742     & 0.726           \\ 
\cline{2-15}
                     & \multirow{3}{*}{200} & 0     & 0.024 & 0.014     & 0.012          & 0.040 & 0.018     & 0.030          & 0.018 & 0.010     & 0.020          & 0.036 & 0.022     & 0.022           \\
                     &                      & 0.2   & 0.644 & 0.646     & 0.594          & 0.560 & 0.542     & 0.500          & 0.646 & 0.588     & 0.566          & 0.626 & 0.614     & 0.576           \\
                     &                      & 0.4   & 0.998 & 0.996     & 0.996          & 0.994 & 0.996     & 0.994          & 0.998 & 0.998     & 0.998          & 1.000 & 1.000     & 0.998           \\ 
\cline{2-15}
                     & \multirow{3}{*}{400} & 0     & 0.024 & 0.024     & 0.020          & 0.050 & 0.034     & 0.034          & 0.056 & 0.034     & 0.036          & 0.036 & 0.014     & 0.012           \\
                     &                      & 0.2   & 0.886 & 0.914     & 0.904          & 0.942 & 0.936     & 0.904          & 0.888 & 0.920     & 0.898          & 0.936 & 0.930     & 0.898           \\
                     &                      & 0.4   & 1.000 & 1.000     & 1.000          & 1.000 & 1.000     & 1.000          & 1.000 & 1.000     & 1.000          & 1.000 & 1.000     & 1.000           \\
\hline
\end{tabular}
\end{adjustbox}
\caption{\label{table:pairedDistribution40}Simulation results for the comparison of quantile curves (range of $\tau$-values) for two paired samples for percentage of censoring equal to 40\%.  The sample size is $n=100,200,400$, the significance level is $\alpha=0.05$ and Diff $=\beta_{21}-\beta_{11}$, so the null hypothesis is satisfied for Diff $=0$. The covariance between the errors is indicated by $\eta$, see \eqref{eq:dependent}.  The abbreviations  Bonf, $T_{L_2}$ and $T_{L_\infty}$ stand for the test statistics defined in (\ref{Bonf}), (\ref{L2}) and (\ref{Linfty}), respectively.}
\end{table}

\begin{table}[H]
\centering
\begin{adjustbox}{totalheight=\textheight-3\baselineskip}
\small
\begin{tabular}{|c|c|c|ccc|ccc|ccc|ccc|} 
\cline{4-15}
\multicolumn{1}{c}{} & \multicolumn{1}{c}{} &       & Bonf  & $T_{L_2}$ & $T_{L_\infty}$ & Bonf  & $T_{L_2}$ & $T_{L_\infty}$ & Bonf  & $T_{L_2}$ & $T_{L_\infty}$ & Bonf  & $T_{L_2}$ & $T_{L_\infty}$  \\ 
\cline{4-15}
\multicolumn{1}{c}{} & \multicolumn{1}{c}{} &       & \multicolumn{3}{c|}{\cite{debacker2019}}     & \multicolumn{3}{c|}{\cite{peng2008survival}}     & \multicolumn{3}{c|}{\cite{portnoy2003censored}}       & \multicolumn{3}{c|}{\cite{wang2009locally}}       \\ 
\hline
Mod.                 & $n$                    & Diff. & \multicolumn{12}{c|}{$\eta = 0.2$}                                                                                                                 \\ 
\hline
\multirow{9}{*}{1}   & \multirow{3}{*}{100} & 0     & 0.010 & 0.000     & 0.010          & 0.012 & 0.006     & 0.012          & 0.004 & 0.008     & 0.004          & 0.006 & 0.002     & 0.006           \\
                     &                      & 0.2   & 0.038 & 0.040     & 0.034          & 0.012 & 0.008     & 0.010          & 0.016 & 0.012     & 0.014          & 0.028 & 0.030     & 0.028           \\
                     &                      & 0.4   & 0.086 & 0.102     & 0.088          & 0.025 & 0.046     & 0.042          & 0.052 & 0.073     & 0.058          & 0.042 & 0.070     & 0.046           \\ 
\cline{2-15}
                     & \multirow{3}{*}{200} & 0     & 0.020 & 0.018     & 0.026          & 0.016 & 0.022     & 0.022          & 0.026 & 0.020     & 0.018          & 0.016 & 0.018     & 0.016           \\
                     &                      & 0.2   & 0.042 & 0.044     & 0.054          & 0.024 & 0.022     & 0.026          & 0.101 & 0.064     & 0.091          & 0.048 & 0.048     & 0.058           \\
                     &                      & 0.4   & 0.216 & 0.238     & 0.198          & 0.157 & 0.131     & 0.141          & 0.243 & 0.225     & 0.211          & 0.170 & 0.214     & 0.162           \\ 
\cline{2-15}
                     & \multirow{3}{*}{400} & 0     & 0.062 & 0.038     & 0.046          & 0.012 & 0.018     & 0.010          & 0.022 & 0.026     & 0.014          & 0.042 & 0.028     & 0.036           \\
                     &                      & 0.2   & 0.092 & 0.090     & 0.088          & 0.110 & 0.132     & 0.102          & 0.078 & 0.086     & 0.072          & 0.126 & 0.154     & 0.146           \\
                     &                      & 0.4   & 0.496 & 0.586     & 0.486          & 0.466 & 0.518     & 0.456          & 0.544 & 0.606     & 0.540          & 0.410 & 0.506     & 0.448           \\ 
\hline
\multirow{9}{*}{2}   & \multirow{3}{*}{100} & 0     & 0.014 & 0.008     & 0.016          & 0.012 & 0.006     & 0.014          & 0.018 & 0.016     & 0.016          & 0.012 & 0.006     & 0.012           \\
                     &                      & 0.2   & 0.050 & 0.048     & 0.052          & 0.032 & 0.032     & 0.045          & 0.028 & 0.024     & 0.030          & 0.040 & 0.042     & 0.040           \\
                     &                      & 0.4   & 0.128 & 0.166     & 0.122          & 0.090 & 0.157     & 0.136          & 0.166 & 0.186     & 0.180          & 0.140 & 0.194     & 0.158           \\ 
\cline{2-15}
                     & \multirow{3}{*}{200} & 0     & 0.026 & 0.014     & 0.022          & 0.028 & 0.012     & 0.030          & 0.020 & 0.028     & 0.014          & 0.024 & 0.032     & 0.024           \\
                     &                      & 0.2   & 0.100 & 0.138     & 0.106          & 0.122 & 0.130     & 0.120          & 0.090 & 0.102     & 0.084          & 0.126 & 0.122     & 0.130           \\
                     &                      & 0.4   & 0.384 & 0.488     & 0.418          & 0.444 & 0.478     & 0.448          & 0.354 & 0.444     & 0.352          & 0.504 & 0.550     & 0.474           \\ 
\cline{2-15}
                     & \multirow{3}{*}{400} & 0     & 0.022 & 0.032     & 0.030          & 0.036 & 0.040     & 0.038          & 0.026 & 0.026     & 0.022          & 0.024 & 0.030     & 0.030           \\
                     &                      & 0.2   & 0.234 & 0.260     & 0.220          & 0.276 & 0.290     & 0.276          & 0.268 & 0.322     & 0.260          & 0.260 & 0.360     & 0.254           \\
                     &                      & 0.4   & 0.806 & 0.900     & 0.846          & 0.834 & 0.904     & 0.842          & 0.902 & 0.948     & 0.892          & 0.882 & 0.950     & 0.888           \\ 
\hline
\multirow{9}{*}{3}   & \multirow{3}{*}{100} & 0     & 0.018 & 0.012     & 0.014          & 0.028 & 0.032     & 0.028          & 0.008 & 0.014     & 0.012          & 0.018 & 0.014     & 0.016           \\
                     &                      & 0.2   & 0.022 & 0.028     & 0.038          & 0.076 & 0.082     & 0.090          & 0.026 & 0.034     & 0.036          & 0.042 & 0.036     & 0.046           \\
                     &                      & 0.4   & 0.360 & 0.496     & 0.392          & 0.204 & 0.270     & 0.208          & 0.306 & 0.396     & 0.348          & 0.326 & 0.384     & 0.370           \\ 
\cline{2-15}
                     & \multirow{3}{*}{200} & 0     & 0.022 & 0.026     & 0.020          & 0.010 & 0.004     & 0.010          & 0.016 & 0.016     & 0.022          & 0.020 & 0.010     & 0.020           \\
                     &                      & 0.2   & 0.190 & 0.282     & 0.202          & 0.176 & 0.216     & 0.206          & 0.296 & 0.288     & 0.274          & 0.194 & 0.264     & 0.202           \\
                     &                      & 0.4   & 0.778 & 0.828     & 0.818          & 0.792 & 0.868     & 0.798          & 0.780 & 0.894     & 0.768          & 0.838 & 0.866     & 0.834           \\ 
\cline{2-15}
                     & \multirow{3}{*}{400} & 0     & 0.020 & 0.022     & 0.024          & 0.022 & 0.012     & 0.018          & 0.020 & 0.016     & 0.022          & 0.016 & 0.020     & 0.018           \\
                     &                      & 0.2   & 0.352 & 0.520     & 0.394          & 0.374 & 0.476     & 0.402          & 0.508 & 0.580     & 0.504          & 0.546 & 0.566     & 0.548           \\
                     &                      & 0.4   & 0.994 & 0.998     & 0.994          & 0.992 & 0.994     & 0.988          & 0.988 & 0.998     & 0.992          & 0.992 & 0.994     & 0.990           \\ 
\hline
\multicolumn{3}{|c|}{}     & \multicolumn{12}{c|}{$\eta = 0.4$}                                                                                                                 \\ 
\hline
\multirow{9}{*}{1}   & \multirow{3}{*}{100} & 0     & 0.032 & 0.022     & 0.032          & 0.006 & 0.016     & 0.010          & 0.016 & 0.008     & 0.016          & 0.016 & 0.006     & 0.012           \\
                     &                      & 0.2   & 0.026 & 0.022     & 0.024          & 0.004 & 0.016     & 0.006          & 0.017 & 0.012     & 0.017          & 0.020 & 0.018     & 0.022           \\
                     &                      & 0.4   & 0.034 & 0.060     & 0.068          & 0.050 & 0.033     & 0.042          & 0.039 & 0.039     & 0.028          & 0.058 & 0.046     & 0.056           \\ 
\cline{2-15}
                     & \multirow{3}{*}{200} & 0     & 0.010 & 0.012     & 0.010          & 0.016 & 0.006     & 0.008          & 0.018 & 0.012     & 0.016          & 0.004 & 0.000     & 0.006           \\
                     &                      & 0.2   & 0.078 & 0.076     & 0.076          & 0.060 & 0.058     & 0.072          & 0.020 & 0.034     & 0.026          & 0.032 & 0.046     & 0.036           \\
                     &                      & 0.4   & 0.254 & 0.256     & 0.232          & 0.175 & 0.236     & 0.171          & 0.233 & 0.337     & 0.275          & 0.190 & 0.246     & 0.168           \\ 
\cline{2-15}
                     & \multirow{3}{*}{400} & 0     & 0.022 & 0.022     & 0.022          & 0.032 & 0.030     & 0.030          & 0.024 & 0.016     & 0.022          & 0.024 & 0.032     & 0.020           \\
                     &                      & 0.2   & 0.168 & 0.206     & 0.170          & 0.104 & 0.116     & 0.136          & 0.080 & 0.082     & 0.084          & 0.140 & 0.154     & 0.144           \\
                     &                      & 0.4   & 0.528 & 0.676     & 0.582          & 0.532 & 0.612     & 0.518          & 0.462 & 0.632     & 0.494          & 0.506 & 0.576     & 0.504           \\ 
\hline
\multirow{9}{*}{2}   & \multirow{3}{*}{100} & 0     & 0.012 & 0.010     & 0.016          & 0.008 & 0.004     & 0.006          & 0.006 & 0.010     & 0.006          & 0.008 & 0.004     & 0.008           \\
                     &                      & 0.2   & 0.066 & 0.066     & 0.060          & 0.026 & 0.049     & 0.030          & 0.040 & 0.051     & 0.047          & 0.030 & 0.036     & 0.030           \\
                     &                      & 0.4   & 0.202 & 0.232     & 0.188          & 0.147 & 0.183     & 0.139          & 0.162 & 0.193     & 0.162          & 0.124 & 0.160     & 0.126           \\ 
\cline{2-15}
                     & \multirow{3}{*}{200} & 0     & 0.026 & 0.016     & 0.022          & 0.012 & 0.014     & 0.016          & 0.016 & 0.018     & 0.014          & 0.012 & 0.010     & 0.010           \\
                     &                      & 0.2   & 0.136 & 0.142     & 0.120          & 0.108 & 0.128     & 0.106          & 0.080 & 0.100     & 0.088          & 0.092 & 0.094     & 0.108           \\
                     &                      & 0.4   & 0.594 & 0.732     & 0.644          & 0.504 & 0.660     & 0.494          & 0.654 & 0.732     & 0.676          & 0.498 & 0.674     & 0.506           \\ 
\cline{2-15}
                     & \multirow{3}{*}{400} & 0     & 0.028 & 0.028     & 0.026          & 0.048 & 0.030     & 0.052          & 0.026 & 0.016     & 0.018          & 0.020 & 0.016     & 0.020           \\
                     &                      & 0.2   & 0.338 & 0.384     & 0.306          & 0.312 & 0.352     & 0.300          & 0.216 & 0.274     & 0.236          & 0.356 & 0.428     & 0.344           \\
                     &                      & 0.4   & 0.934 & 0.954     & 0.946          & 0.944 & 0.984     & 0.964          & 0.972 & 0.988     & 0.966          & 0.974 & 0.988     & 0.972           \\ 
\hline
\multirow{9}{*}{3}   & \multirow{3}{*}{100} & 0     & 0.014 & 0.000     & 0.014          & 0.012 & 0.006     & 0.014          & 0.004 & 0.000     & 0.006          & 0.008 & 0.002     & 0.010           \\
                     &                      & 0.2   & 0.078 & 0.062     & 0.088          & 0.104 & 0.062     & 0.110          & 0.036 & 0.054     & 0.050          & 0.040 & 0.084     & 0.066           \\
                     &                      & 0.4   & 0.454 & 0.548     & 0.472          & 0.325 & 0.431     & 0.341          & 0.324 & 0.439     & 0.360          & 0.320 & 0.444     & 0.364           \\ 
\cline{2-15}
                     & \multirow{3}{*}{200} & 0     & 0.010 & 0.008     & 0.012          & 0.032 & 0.012     & 0.022          & 0.012 & 0.012     & 0.016          & 0.010 & 0.006     & 0.008           \\
                     &                      & 0.2   & 0.326 & 0.352     & 0.320          & 0.284 & 0.320     & 0.284          & 0.240 & 0.332     & 0.216          & 0.206 & 0.214     & 0.210           \\
                     &                      & 0.4   & 0.882 & 0.914     & 0.892          & 0.872 & 0.916     & 0.848          & 0.818 & 0.924     & 0.826          & 0.716 & 0.858     & 0.730           \\ 
\cline{2-15}
                     & \multirow{3}{*}{400} & 0     & 0.008 & 0.008     & 0.008          & 0.016 & 0.016     & 0.014          & 0.020 & 0.016     & 0.020          & 0.014 & 0.008     & 0.016           \\
                     &                      & 0.2   & 0.622 & 0.714     & 0.618          & 0.578 & 0.668     & 0.502          & 0.708 & 0.782     & 0.700          & 0.446 & 0.618     & 0.432           \\
                     &                      & 0.4   & 0.996 & 1.000     & 0.996          & 0.996 & 0.998     & 0.996          & 1.000 & 1.000     & 1.000          & 1.000 & 1.000     & 1.000           \\
\hline
\end{tabular}
\end{adjustbox}
\caption{\label{table:pairedMedian40}Simulation results for the comparison of the median ($\tau=0.5$) for two paired samples subject to $40\%$ censoring. The sample size is $n=100,200,400$, the significance level is $\alpha=0.05$ and Diff $=\beta_{21}-\beta_{11}$, so the null hypothesis is satisfied for Diff $=0$. The covariance between the errors is indicated by $\eta$, see \eqref{eq:dependent}. The abbreviations Bonf, $T_{L_2}$ and $T_{L_\infty}$ stand for the test statistics defined in (\ref{Bonf}), (\ref{L2}) and (\ref{Linfty}), respectively. }
\end{table}

\end{document}